\documentclass[aps,twocolumn,amsmath,amssymb,floatfix,superscriptaddress]{revtex4}
\usepackage{graphicx}
\usepackage{dcolumn}
\usepackage{bm}
\usepackage{amsfonts}
\usepackage{graphicx}% Include figure files
\usepackage{dcolumn}% Align table columns on decimal point
\usepackage{bm}% bold math
\usepackage{amsmath}% needed for subequations
\usepackage{amssymb}
\usepackage{gensymb}
\usepackage{braket}
\usepackage{multirow}
\usepackage{ctable}
\usepackage{hyperref}

\begin{document}
\title{Optical conductivity of black phosphorus with a tunable electronic structure}
\author{Jiho Jang}
\affiliation{Department of Physics and Astronomy, Seoul National University, Seoul 08826, Korea}
\author{Seongjin Ahn}
\affiliation{Department of Physics and Astronomy, Seoul National University, Seoul 08826, Korea}
\affiliation{Center for Correlated Electron Systems, Institute for Basic Science (IBS), Seoul 08826, Korea}
\author{Hongki Min}
\affiliation{Department of Physics and Astronomy, Seoul National University, Seoul 08826, Korea}
\email{hmin@snu.ac.kr}
\date{\today}

\begin{abstract}
%We study the optical conductivity of a biased few-layer black phosphorus focusing on the low-energy characteristic frequency dependence in each phase. ...
Black phosphorus (BP) is a two-dimensional layered material composed of phosphorus atoms. Recently, it was demonstrated that external perturbations such as an electric field close the band gap in few-layer BP, and can even induce a band inversion, resulting in an insulator phase with a finite energy gap or a Dirac semimetal phase characterized by two separate Dirac nodes. At the transition between the two phases, a semi-Dirac state appears in which energy disperses linearly along one direction and quadratically along the other. In this work, we study the optical conductivity of few-layer BP using a lattice model and the corresponding continuum model, incorporating the effects of an external electric field and finite temperature. We find that the low-frequency optical conductivity scales a power law that differs depending on the phase, which can be utilized as an experimental signature of few-layer BP in different phases. We also systematically analyze the evolution of the material parameters as the electric field increases, and the consequence on the power-law behavior of the optical conductivity.
\end{abstract}

\maketitle
%\normalsize

%%%%%%%%%%%%%%%%%%%%%%%%%%%%%%%%%%%%%%%%%%%%%%%%%%%%%%%%%%%%%%%%%%%%%%%%%%%%%%%%%%%%%%%%%%%%%%%%%%%%
\section{Motivation}
\begin{figure}[htb]
\includegraphics[width=1.0\linewidth]{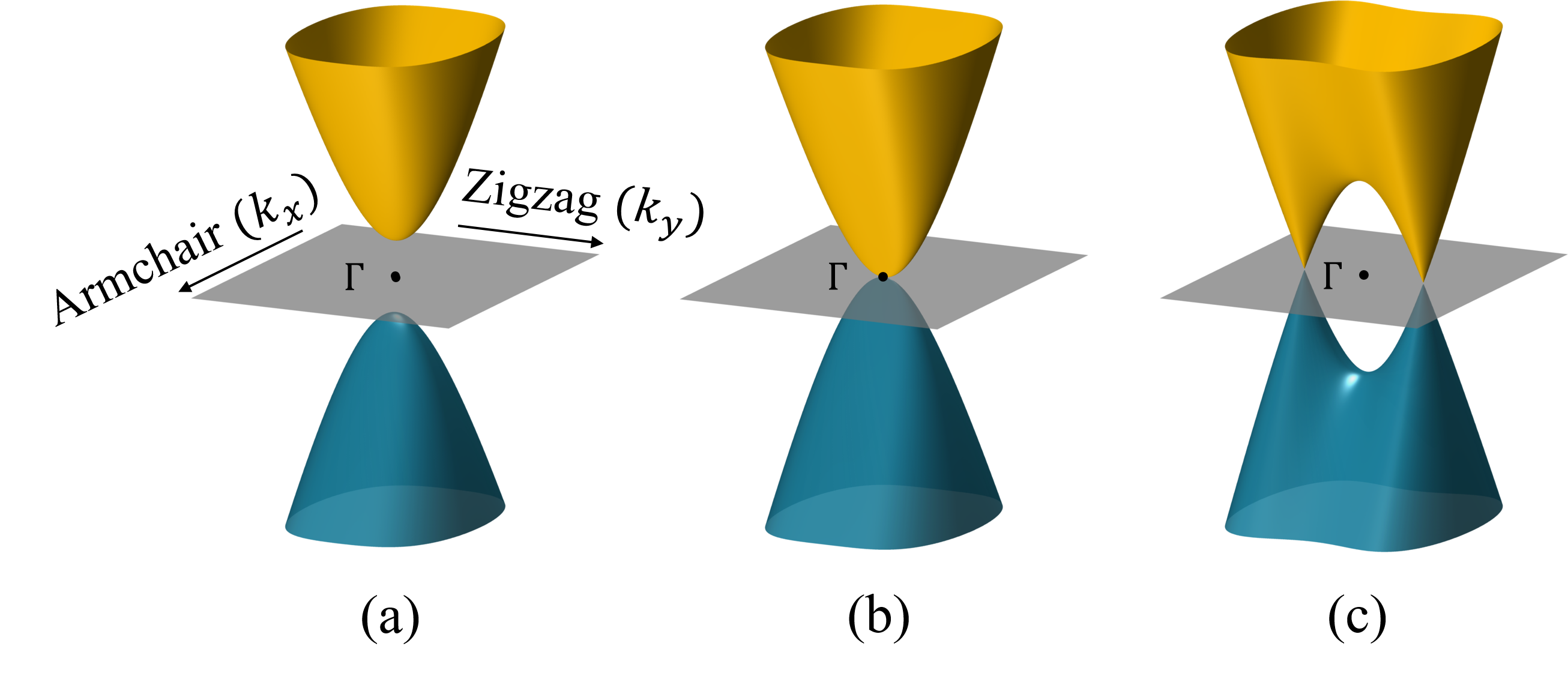}
\caption{
Low-energy band structure of few-layer BP in the (a) insulator phase, (b) semi-Dirac point, and (c) Dirac semimetal phase.
}
\label{fig:electronic_structure}
\end{figure}

Black phosphorus (BP) is a two-dimensional (2D) layered material composed of phosphorus atoms, where the layers are stabilized by weak van der Waals forces, and thus can be exfoliated into a few-layer form. (For a recent review, see \cite{chaves2017theoretical}.) It is known that the band gap of BP decreases as the thickness increases from ~1.6 eV for a monolayer to ~0.3 eV in bulk \cite{li2017direct, qiao2014high, tran2014layer}. Recently, it was discovered that external perturbations such as pressure \cite{PhysRevB.91.195319, PhysRevLett.115.186403}, strain \cite{rodin2014strain}, an electric field  \cite{liu2015switching, PhysRevB.93.245433, doh2017dirac} and surface doping \cite{kim2015observation, baik2015emergence, kim2017two} close the band gap, and can even induce band inversion in few-layer BP. This results in three different states in few-layer BP: an insulator phase with a gap, a semi-Dirac point with gapless anisotropic dispersion (linear and quadratic in the armchair and zigzag directions, respectively), and a Dirac semimetal phase with two Dirac points, as illustrated in Fig.~\ref{fig:electronic_structure}.

%%%%%%%%%%%%%%%%%%%%%%%%%%%%%%%%%%%%%%%%%%%%%%%%%%
%%%%%%%%%%%%%%%%%%%%%%%%%%%%%%%%%%%%%%%%%%%%%%%%%%

There have been a significant number of theoretical and experimental studies concerning various physical properties of few-layer BP, including the electronic structure \cite{li2014black, li2014electrons, han2014electronic}, optical properties \cite{low2014tunable, xia2014rediscovering, yuan2015transport, mao2015optical, yuan2015polarization,  lin2016multilayer, doi:10.1021/acs.nanolett.6b03362, doi:10.1021/acs.nanolett.7b03050, torbatian2018optical}, transport properties \cite{yuan2015transport, doganov2015transport, PhysRevB.93.125113, liu2017temperature, park2018semiclassical}, and Landau levels \cite{pereira2015landau, tahir2015magneto, PhysRevB.93.245433, PhysRevB.92.165405, Zhou2015}. The tight-binding model for few-layer BP has been proposed by several groups \cite{rudenko2014quasiparticle, PhysRevB.92.085419, de2017multilayered}.

However, to our knowledge there has been no systematic study on the optical conductivity of few-layer BP in each phase and the corresponding characteristic frequency dependence. In this study, we investigate the optical conductivity of few-layer BP with AB stacking type, which is the most common and energetically stable stacking configuration \cite{ccakir2015significant, wu2015atomic}. We conduct both numerical and analytical calculations using a lattice model and the corresponding continuum model, which contain the two phases and capture all the low-energy optical properties.
%with a model Hamiltonian which contains the two phases and confirm that the minimal continuum model can capture all the essential low-energy optical properties.
As the perpendicular external electric field increases, the self-consistently obtained energy gap parameter $\varepsilon_{\rm g}$ changes from a positive to a negative value, and the optical conductivity exhibits a characteristic frequency dependence in each phase for both armchair and zigzag directions, which can be utilized optically to identify each phase in few-layer BP.
%Especially, the gapless anisotropic dispersion in the semi-Dirac point has important consequences for the low-frequency optical properties which can be used as an experimental fingerprint for the existence of an anisotropic Dirac point.

This paper is organized as follows. In Sec.~\ref{sec:model}, we introduce the tight-binding lattice model Hamiltonian and continuum model Hamiltonian for few-layer BP. In Sec.~\ref{sec:screening_theory}, we study the influence of an external electric field on few-layer BP within a mean-field Hartree approximation, and self-consistently obtain the electronic band structure, as well as the evolution of the model parameters with the external electric field. In Sec.~\ref{sec:optical conductivity}, we present the results calculated for the optical conductivity based on the two models along with the analytic results for each phase. We also describe the effect of a finite temperature on the optical conductivity, demonstrating a power-law change at low frequencies. Finally, in Sec~\ref{sec:discussion} we conclude with a discussion on the effect of the number of layers and the intraband response.
%%%%%%%%%%%%%%%%%%%%%%%%%%%%%%%%%%%%%%%%%%%%%%%%%%%%%%%%%%%%%%%%%%%%%%%%%%%%%%%%%%%%%%%%%%%%%%%%%%%%

\section{Model}
\label{sec:model}
\subsection{Lattice model}
\label{subsec:lattice_model}
\begin{figure}[!htb]
\includegraphics[width=1.0\linewidth]{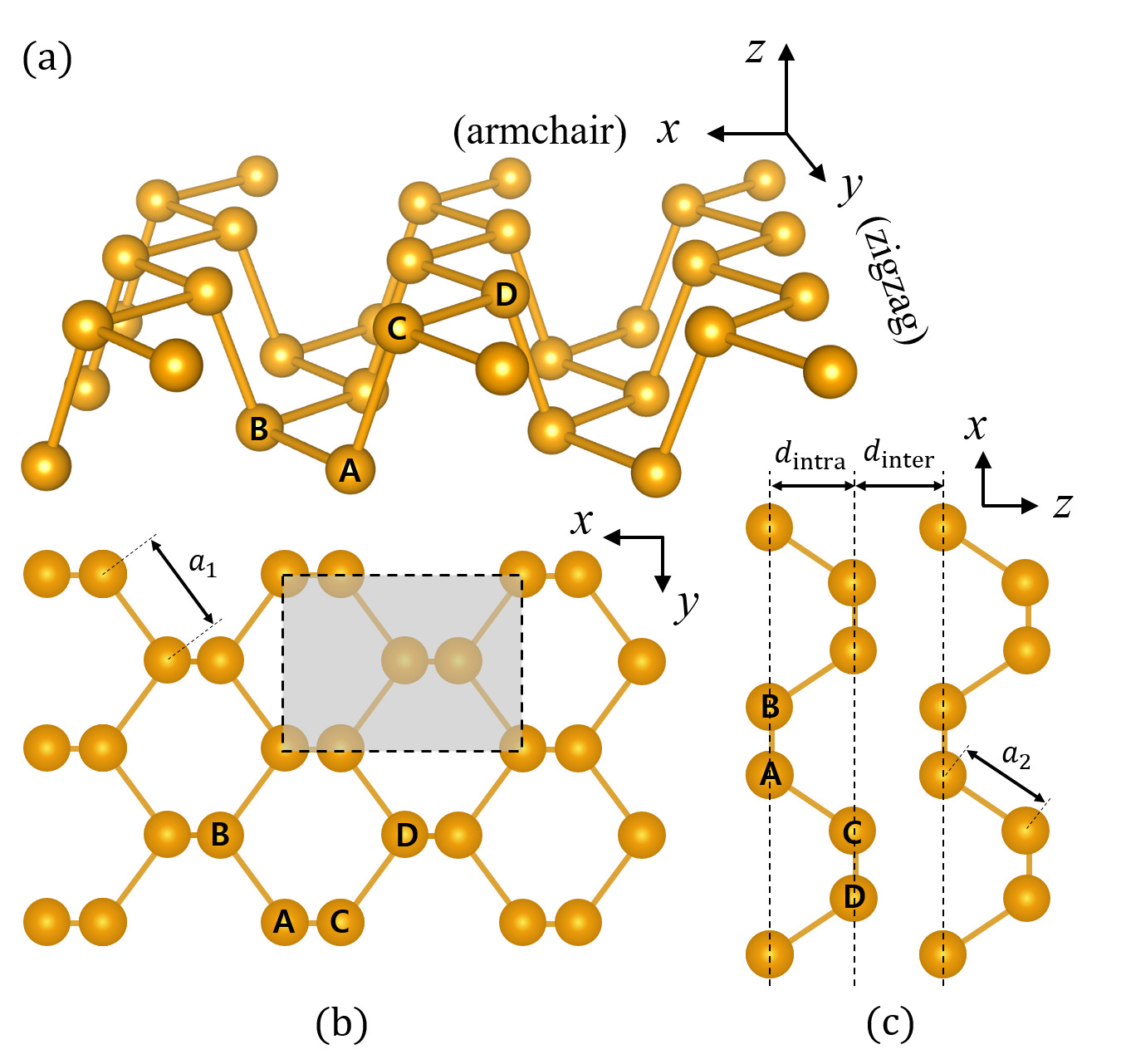}
\caption{
(a) Atomic structure of monolayer BP and (b) its top view. (c) Side view of bilayer BP. Here, the $x$ and $y$ axes are set along the armchair and zigzag directions, respectively. The shaded rectangle in (b) indicates the unit cell of BP. The interatomic distances $a_1$ and $a_2$ are given by $a_1 = 2.21 {\rm \AA}$ and $a_2 = 2.24 {\rm \AA}$, and the interlayer and intralayer distances are given by $d_{\rm inter} = 3.17 {\rm \AA}$ and $d_{\rm intra} = 2.13 {\rm \AA}$, respectively.
}
\label{fig:atomic_structure}
\end{figure}

In this section, we introduce a tight-binding model for few-layer BP, along with its crystal structure shown in Fig.~\ref{fig:atomic_structure}. Few-layer BP exhibits a buckled honeycomb lattice structure, with four phosphorus atoms in each unit cell. The tight-binding lattice model for few-layer BP in the basis of sublattices with $3s$, $3p_x$, and $3p_z$ orbitals is given by
\begin{eqnarray}
\label{eq:lattice_model}
H \!\!&=&\!\! \sum_{l,i} \varepsilon_{l,i} c_{l,i}^\dagger c_{l,i} + \!\sum_{l,i\neq j} t_{i,j}^{l} c_{l,i}^\dagger c_{l,j}+\!\!\!\sum_{l\neq m,i,j}\!\! t_{i,j}^{l,m} c_{l,i}^\dagger c_{m,j}
\end{eqnarray}
where $c_{l,i}^{\dagger}$ ($c_{l,i}$) corresponds to the creation (annihilation) operator for an electron on the $i$th site in the $l$th layer, and $t_{i,j}^{l}$ and $t_{i,j}^{l,m}$ are intralayer and interlayer hopping parameters, respectively. Here, $\varepsilon_{l,i}$ is the on-site energy,
%which is tuned by applying an external perpendicular field and determines the phase of the system between insulator and Dirac semimetal phases, as shown in Fig.~\ref{fig:electronic_structure}.
which is self-consistently determined in the presence of an external perpendicular electric field, as will be discussed in Sec.~\ref{sec:screening_theory}.
In this work, we consider intralayer hopping terms for up to 10 nearest neighbors, and the five nearest-neighbor interlayer hopping terms given in~\cite{PhysRevB.92.085419}.
%%%%%%%%%%%%%%%%%%%%%%%%%%%%%%%%%%%%%%%%%%%%%%%%%%
%%%%%%%%%%%%%%%%%%%%%%%%%%%%%%%%%%%%%%%%%%%%%%%%%%
\subsection{Continuum model}
\label{subsec:continuum model}
The low-energy expansion of the tight-binding Hamiltonian in Eq. (\ref{eq:lattice_model}) around the $\Gamma$ point yields
%Next, we construct the low-energy effective Hamiltonian for a biased few-layer BP which can describe the low-energy regime of the tight-binding Hamiltonian in Eq.~(\ref{eq:lattice_model}) and it is given by
\begin{equation}
\label{eq:continuum_model_1}
H=\left(
\begin{array}{cc}
\varepsilon_{\rm CB} + a_x k_x^2 + a_y k_y^2 & -i t k_x \\
i t^{\ast} k_x & \varepsilon_{\rm VB}  + b_x k_x^2 + b_y k_y^2
\end{array}
\right),
\end{equation}
where $\varepsilon_{\rm CB}$ is the conduction band minimum and $\varepsilon_{\rm VB}$ is the valence band maximum. Note that linear terms in $k_y$ are not allowed in the off-diagonal element of the Hamiltonian, owing to the reflection symmetry with respect to the $y=0$ plane ($\mathcal{M}_y$) \cite{rodin2014strain, kim2017two}. From the energy dispersion of the tight-binding model, we confirm that we can effectively set $a_x \approx -b_x$ and $a_y \approx -b_y$ \cite{baik2015emergence}, leading to
\begin{equation}
\label{eq:continuum_model_2}
H=\hbar v k_x \sigma_y + \left(\frac{1}{2}\varepsilon_{\rm g}+\gamma{\hbar^2 k_x^2\over 2m}+{\hbar^2 k_y^2\over 2m}\right) \sigma_z,
\end{equation}
where $\varepsilon_{\rm g}=\varepsilon_{\rm CB}-\varepsilon_{\rm VB}$, $v$ is the effective velocity along the armchair direction, and $m$ is the effective mass along the zigzag direction. Here, we set $\frac{1}{2}(\varepsilon_{\rm CB}+\varepsilon_{\rm VB})=0$ to be the zero of the energy.

Note that the parabolic term $\gamma {\hbar^2 k_x^2\over 2m}\sigma_z$, whose contribution is characterized by the dimensionless parameter $\gamma$, is added in the low-energy continuum model in the armchair direction beyond the lowest-order linear term $\hbar v k_x \sigma_y$. We included this term to take into account its role in the optical conductivity, especially at high frequencies, as will be discussed in Sec.~\ref{sec:optical conductivity}.

Figure \ref{fig:electronic_structure} illustrates the energy dispersions depending on the sign of $\varepsilon_{\rm g}$.
When $\varepsilon_{\rm g}>0$, the system is in the insulator phase, and $\varepsilon_{\rm g}$ corresponds to the size of the energy gap [Fig.~\ref{fig:electronic_structure}(a)].
When the band gap closes ($\varepsilon_{\rm g}=0$), the system is described by a 2D semi-Dirac Hamiltonian \cite{banerjee2012phenomenology, sriluckshmy2018interplay} [Fig.~\ref{fig:electronic_structure}(b)], where the energy dispersion is linear along the armchair direction ($k_x$) and quadratic along the zigzag direction ($k_y$). When $\varepsilon_{\rm g}<0$, a band inversion occurs, and the semi-Dirac point splits into two separated Dirac points located at ${\bm k} = (0, \pm \sqrt{\frac{m |\varepsilon_{\rm g}|}{\hbar^2}})$ [Fig.~\ref{fig:electronic_structure}(c)].
%%%%%%%%%%%%%%%%%%%%%%%%%%%%%%%%%%%%%%%%%%%%%%%%%%%%%%%%%%%%%%%%%%%%%%%%%%%%%%%%%%%%%%%%%%%%%%%%%%%%

\section{Screening Theory}
\label{sec:screening_theory}
In this section, we explore the relation between the band structure of few-layer BP and its dual-gate configuration within a self-consistent Hartree approximation. We consider the situation in which few-layer BP is located between the two metallic gates, whose charge densities are given by $n_{\rm tg}\le 0$ (top gate) and $n_{\rm bg}\ge 0$ (bottom gate). By tuning the gate voltages, one can manipulate both the electric field applied to the few-layer BP and the gate-induced charge density in each layer of BP. In the following, we explain the self-consistent Hartree formalism for few-layer BP, and present numerical results obtained by solving the self-consistent Hartree equation.

\subsection{Self-consistent Hartree approximation}
We begin with the non-interacting Hamiltonian for a layered system,
\begin{eqnarray}
H_0 &=& \sum_{{\bm k}, \lambda, \lambda'} \varepsilon^{(0)}_{\lambda, \lambda'}\left({\bm k}\right) c^\dagger_{{\bm k}, \lambda} c_{{\bm k}, \lambda'},
\end{eqnarray}
where $c^\dagger_{{\bm k}, \lambda}(c_{{\bm k}, \lambda})$ are creation (annihilation) operators for the wave vector ${\bm k}$ and state $\lambda$ (including spin, orbital and layer degrees of freedom). Next, we incorporate the electron-electron Coulomb interaction given by
\begin{eqnarray}
V &=& \frac{1}{2} \sum_{{\bm k}, {\bm k'}, {\bm q}} \sum_{\lambda, \lambda'} \widetilde{V}_{\lambda \lambda'}({\bm q}) c^\dagger_{{\bm k}+{\bm q}, \lambda} c^\dagger_{{\bm k'}-{\bm q}, \lambda'} c_{{\bm k'}, \lambda'} c_{{\bm k}, \lambda},
\end{eqnarray}
where $\widetilde{V}_{\lambda \lambda'}({\bm q})= \frac{2\pi e^2}{\epsilon |{\bm q}|}e^{-|{\bm q}| d_{\lambda \lambda'}}$ is the 2D Fourier transform of the real-space Coulomb interaction $\widetilde{V}_{\lambda \lambda'}({\bm x})=\frac{e^2}{\epsilon\sqrt{|x|^2+d_{\lambda \lambda'}^2}}$, and $d_{\lambda \lambda'}$ refers to the distance between the $\lambda$ and $\lambda'$ states.

By employing a mean-field Hartree approximation, we can reduce the full Hamiltonian $H = H_0 + V$ to
\begin{align}
H_{\rm MF} = H_0 + \sum_{{\bm k}, \lambda} \varepsilon^{({\rm H})}_{\lambda} c^\dagger_{{\bm k}, \lambda} c_{{\bm k}, \lambda},
\end{align}
where
\begin{align}
\varepsilon^{({\rm H})}_{\lambda} &= \sum_{\lambda'} \widetilde{V}_{\lambda \lambda'}(0) n_{\lambda'},
%\braket{V^{({\rm H})}} &= \frac{1}{2} \sum_{{\bm k}, \lambda}\varepsilon^{(H)}_{\lambda} n_{\lambda}({\bm k}).
\end{align}
and $n_{\lambda}=\sum_{\bm k}\braket{c^\dagger_{{\bm k} \lambda}c_{{\bm k} \lambda}}$ \cite{min2007ab}. The induced potential difference between the $\lambda$ and $\lambda'$ states is given by
\begin{align}
\varepsilon^{({\rm H})}_{\lambda}-\varepsilon^{({\rm H})}_{\lambda'} &= \sum_{\lambda''} \left[\widetilde{V}_{\lambda \lambda''}(0)-\widetilde{V}_{\lambda' \lambda''}(0)\right] n_{\lambda''} \nonumber \\
&=\sum_{\lambda''} \frac{2\pi e^2}{\epsilon} \left(d_{\lambda' \lambda''} - d_{\lambda \lambda''}\right) n_{\lambda''}.
\label{eq:potential_difference}
\end{align}
Note that $\widetilde{V}_{\lambda \lambda''}(0)-\widetilde{V}_{\lambda' \lambda''}(0)=\frac{2\pi e^2}{\epsilon} \left(d_{\lambda' \lambda''} - d_{\lambda \lambda''}\right)$ can be obtained by taking the limit $|\bm{q}|\rightarrow 0$.
%{\bf [Min: Alternatively, since the equation below is basic, we can simply mention that ``Note that $\widetilde{V}_{\lambda \lambda''}(0)-\widetilde{V}_{\lambda' \lambda''}(0)=\frac{2\pi e^2}{\epsilon} \left(d_{\lambda' \lambda''} - d_{\lambda \lambda''}\right)$ can be obtained by taking the limit $|\bm{q}|\rightarrow 0$."]}
Therefore, the total onsite energy difference between the $\lambda$ and $\lambda'$ states including the contribution from the Hartree potential and that from the external electric field $E_{\rm ext}$, is given by
% \begin{align}
% \varepsilon^{({\rm tot})}_{\lambda} &- \varepsilon^{({\rm tot})}_{\lambda'} \nonumber \\
% &= \left(\varepsilon^{({\rm ext})}_{\lambda} - \varepsilon^{({\rm ext})}_{\lambda'}\right) + \left(\varepsilon^{({\rm H})}_{\lambda} - \varepsilon^{({\rm H})}_{\lambda'} \right) \nonumber \\
% &= e E_{\rm ext} d_{\lambda \lambda'} +\sum_{{\bm k}, \lambda''} \frac{2\pi e^2}{\epsilon} \left(d_{\lambda' \lambda''} - d_{\lambda \lambda''}\right) n_{\lambda''}({\bm k}),
% \label{eq:hartree_final}
% \end{align}
%\end{align}
%where $\varepsilon^{({\rm ext})}_{\lambda}$ is the onsite energy (due to ???) an external electric field and $E_{\rm ext}$ is an external electric field strength.
\begin{eqnarray}
\label{eq:hartree_final}
\varepsilon^{({\rm tot})}_{\lambda}- \varepsilon^{({\rm tot})}_{\lambda'} = \varepsilon^{(0)}_{\lambda\lambda}-\varepsilon^{(0)}_{\lambda'\lambda'}+\varepsilon^{({\rm H})}_{\lambda}-\varepsilon^{({\rm H})}_{\lambda'} + e E_{\rm ext} d_{\lambda\lambda'}.
\end{eqnarray}

Because the whole system including the top/bottom gates and the sample in-between is charge-neutral, the sum of the top-gate, bottom-gate, and sample charge densities must be zero. Thus, for the given top-gate ($n_{\rm tg}$) and bottom-gate ($n_{\rm bg}$) charge densities, the total carrier density, $n_{\rm tot} = -(n_{\rm tg}+n_{\rm bg})$ is induced for the sample, and the Fermi energy $\varepsilon_{\rm F}$ of the sample can be calculated from $n_{\rm tot}$. The top-gate and bottom-gate charge densities also determine the external electric field as $E_{\rm ext}=\frac{1}{2}(E_{\rm tg}+E_{\rm bg})$, where $E_{\rm tg}=\frac{4\pi e}{\epsilon}n_{\rm tg}$ and $E_{\rm bg}=-\frac{4\pi e}{\epsilon}n_{\rm bg}$. Therefore, by solving Eqs.~(\ref{eq:potential_difference}) and (\ref{eq:hartree_final}) for the given $n_{\rm tot}$ (or $\varepsilon_{\rm F}$) and $E_{\rm ext}$, the onsite energies $\varepsilon^{({\rm tot})}_{\lambda}$ can be self-consistently obtained.

The approach we adopted above, a self-consistent Hartree method, is essentially equivalent to solving the self-consistent Poisson equation presented in Li \textit{et al.} \cite{li2018tuning}.
When an external electric field is applied, the charge carriers in a few-layer BP system are redistributed in such a manner that the electrostatic energy of the system is minimized. (see~\cite{li2018tuning} for details of charge distribution for various set-ups.) In this sense, the Hartree contribution consists precisely of the classical electrostatic potential generated by induced charges. %The distribution of the charge carriers is also systematically studied in Sec. IIIB in Ref. \cite{li2018tuning} and we confirmed that our calculation yields the equivalent result.

In the next section, we directly apply Eqs.~(\ref{eq:potential_difference}) and (\ref{eq:hartree_final}) to few-layer BP to analyze the influence of an external electric field. For this calculation, we employ the intralayer distance $d_{\rm intra}=2.13$ ${\rm \AA}$ and interlayer distance $d_{\rm inter}=3.17$ ${\rm \AA}$. We set the dielectric constant $\epsilon=1$ for simplicity, and the choice of $\epsilon$ does not change our results qualitatively.
%and dielectric constant $\epsilon=1$, for simplicity since it does not affect our result on the optical conductivity such as power law.
%the following parameters.. where $d_\mathrm{intra}$ is the distance between ... and $d_\mathrm{inter}$... Here $d_\mathrm{inter}$ and $d_\mathrm{intra}$ represent the distance between ...
%Note that there are two sublayers in monolayer BP due to its puckered structure.
%Thus, for fey-layer BP, there are both perpendicular intralayer and interlayer distance, as shown in Fig.~\ref{fig:atomic_structure}. For numerical calculation, we use $d_{intra}=2.13 \textup{\AA}$ and $d_{inter} = 3.17 \textup{\AA}$.

\subsection{Evolution of the material parameters}
\begin{figure}[!h]
\includegraphics[width=1\linewidth]{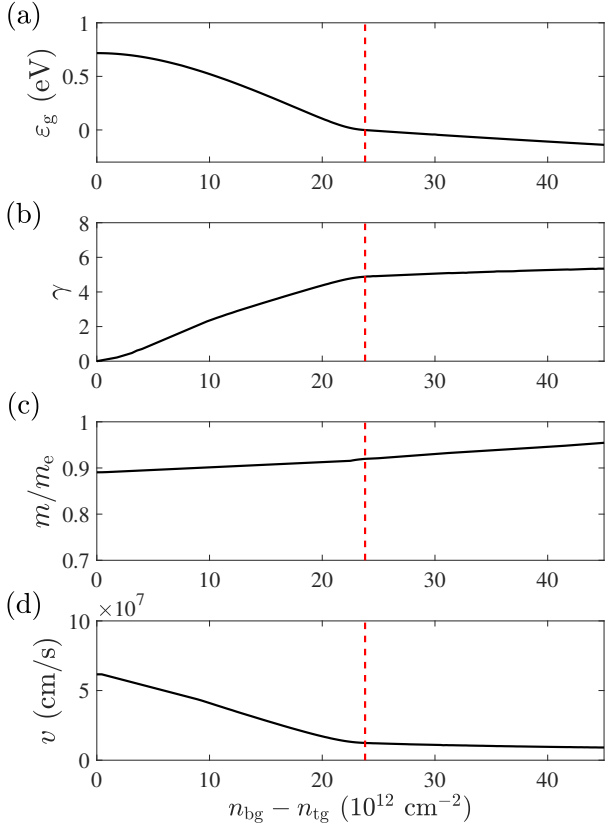}
\caption
{Evolution of the parameters (a) $\varepsilon_{\rm g}$, (b) $\gamma$, (c) $m/m_{\rm e}$, and (d) $v$ as a function of $n_{\rm bg}-n_{\rm tg}$ under the charge neutral condition. Here, $m_{\rm e}$ is the electron mass. The red dotted line represents the semi-Dirac point corresponding to the phase boundary between the insulator phase (left side) and Dirac semimetal phase (right side).
}
\label{fig:parameter_evolution}
\end{figure}

Figure \ref{fig:parameter_evolution} illustrates the evolution of the energy gap parameter $\varepsilon_{\rm g}$, $\gamma$, effective mass $m$, and velocity $v$ under the charge neutral condition ($n_{\rm tg} + n_{\rm tg} = 0$). As the electric field (or equivalently, $n_{\rm bg} - n_{\rm tg}$) increases, $\gamma$ and $m$ increase, whereas $\varepsilon_{\rm g}$ and $v$ decrease. After reaching the semi-Dirac point, the parameters vary slowly with the electric field, owing to enhanced screening.

%%%%%%%%%%%%%%%%%%%%%%%%%%%%%%%%%%%%%%%%%%%%%%%%%%
%%%%%%%%%%%%%%%%%%%%%%%%%%%%%%%%%%%%%%%%%%%%%%%%%%

\section{Optical conductivity}
\label{sec:optical conductivity}
\subsection{Kubo formula}
The Kubo formula for the optical conductivity in the non-interacting and clean limit can be expressed as \cite{mahan2013many}
\begin{align}
\label{eq:conductivity}
\begin{split}
\sigma_{ij}(\omega)
&=- \frac{ie^2}{\hbar} \sum_{s,s'} \int \frac{d^2 k}{(2\pi)^2} \frac{f_{s, \bm{k}}-f_{s',\bm{k}}}{\varepsilon_{s,\bm{k}}-\varepsilon_{s',\bm{k}}}\\ &\times\frac{M^{ss'}_i(\bm k)M^{s's}_j(\bm k)}{\hbar\omega+\varepsilon_{s,\bm{k}}-\varepsilon_{s',\bm{k}}+i0^+},
\end{split}
\end{align}
where $i,j=x,y,z$, $f_{s,\bm{k}}=1/[1+e^{(\varepsilon_{s,\bm{k}}-\mu)/k_{\rm B}T}]$ is the Fermi distribution function for the band index $s$ and wave vector $\bm{k}$, $\mu$ is the chemical potential, and $M^{ss'}_i(\bm k)=\langle{s,\bm{k}}|\hbar\hat{v}_i |{s',\bm{k}'}\rangle$, with the velocity operator $\hat{v}_i$ obtained from the relation $\hat{v}_i=\frac{1}{\hbar}\frac{\partial H}{\partial  k_i}$.
%We find that the transverse optical conductivity $\sigma_{ij}(\omega)$ with $i\neq j$ vanishes in our continuum model, owing to time-reversal symmetry.
In the following, we only consider the real part of the longitudinal optical conductivity in the clean limit.

\subsection{Optical conductivity for each phase}
\label{subsec:optical_conductivity_for_each_phase}
In this section, we present the real part of the optical conductivity of few-layer BP for the lattice model [Eq.~(\ref{eq:lattice_model})] and continuum model [Eq.~(\ref{eq:continuum_model_2})]. For the lattice calculations, we self-consistently obtain the on-site energies in the presence of a perpendicular external electric field, as explained in Sec.~\ref{sec:screening_theory}. Here, we focus on tetralayer BP, and we discuss the effect of the number of layers later in Sec.~\ref{sec:discussion}.
%[{\bf Ahn: I think it would be better if this can be elaborated}].
For the continuum model, we employ a set of parameters obtained by fitting to the lattice model near the $\Gamma$ point, for comparison.
%For the comparison between the lattice and continuum models, we use parameter sets that fit the continuum band structure to the lattice one around the $\Gamma$ point
%(For details of fitting parameters, see the Appendix \ref{app:evolution}).
%In this section, we present the real part of optical conductivity of few-layer BP calculated based on the two-band continuum model in Eq.~(\ref{eq:continuum_model_2_dimensionless}), the lattice model in Eq.~(\ref{eq:lattice_model}) and the analytic results assuming $\gamma = 0$ of the continuum model in Eq.~(\ref{eq:continuum_model_2_dimensionless}). For simplicity, we use tetralayer BP ($N=4$) tight-binding model for the lattice calculation and use paramters obtained from band fitting data of tetralyaer BP to for the continuum model calculation (for details of fitting parameters, see the Appendix \ref{subsec:band_fitting_data}.

\subsubsection{Insulator phase}
\label{subsec:optical_conductivity_normal_insulator}
%For the normal insulator phase ($\varepsilon_{\rm g}>0$), we obtain the leading-order $\omega$ dependence of optical conductivities at zero temperature in the vicinity of $\omega=\varepsilon_{\rm g}/\hbar$:
%%For the normal insulator phase ($\varepsilon_{\rm g}>0$), we obtain the leading-order $\omega$ dependence of optical conductivities at zero temperature in the vicinity of $\omega=\varepsilon_{\rm g}/\hbar$:
%\begin{subequations}
%\begin{eqnarray}
%\nonumber
%\sigma_{xx}(\omega)&\sim& g_{\rm s} \frac{e^2}{4\hbar}
%\sqrt{\frac{\varepsilon_0}{\varepsilon_{\rm g}}}
%\left(
%\frac{1}{2}-\frac{11}{16\varepsilon_{\rm g}}
%\left(\hbar \omega -\varepsilon_{\rm g} \right)
%\right)\\
%&&\times \Theta(\hbar\omega-\varepsilon_{\rm g}), \\
%\nonumber
%\sigma_{yy}(\omega)&\sim& g_{\rm s} \frac{e^2}{4\hbar}
%\left(\frac{\varepsilon_0}{\varepsilon_{\rm g}}\right)^{3/2}
%\left(\frac{\hbar \omega -\varepsilon_{\rm g}}{\varepsilon_0}\right)^2
%\\
%&&\times\Theta(\hbar\omega-\varepsilon_{\rm g}),
%\end{eqnarray}
%\end{subequations}
%where $\varepsilon_0 =  2m_y v_x^2$ and $g_{\rm s}=2$ accounts for spin degeneracy.
%The full analytical results including higher order corrections in $\omega$ can be found in the Appendix \ref{sec:analytic_expressions}.
%%%%%%%%%%%%%%%%%%%%%%%%%%%%%%%%%%%%%%%%%%%%%%%%%%
\begin{figure}[t]
\includegraphics[width=1\linewidth]{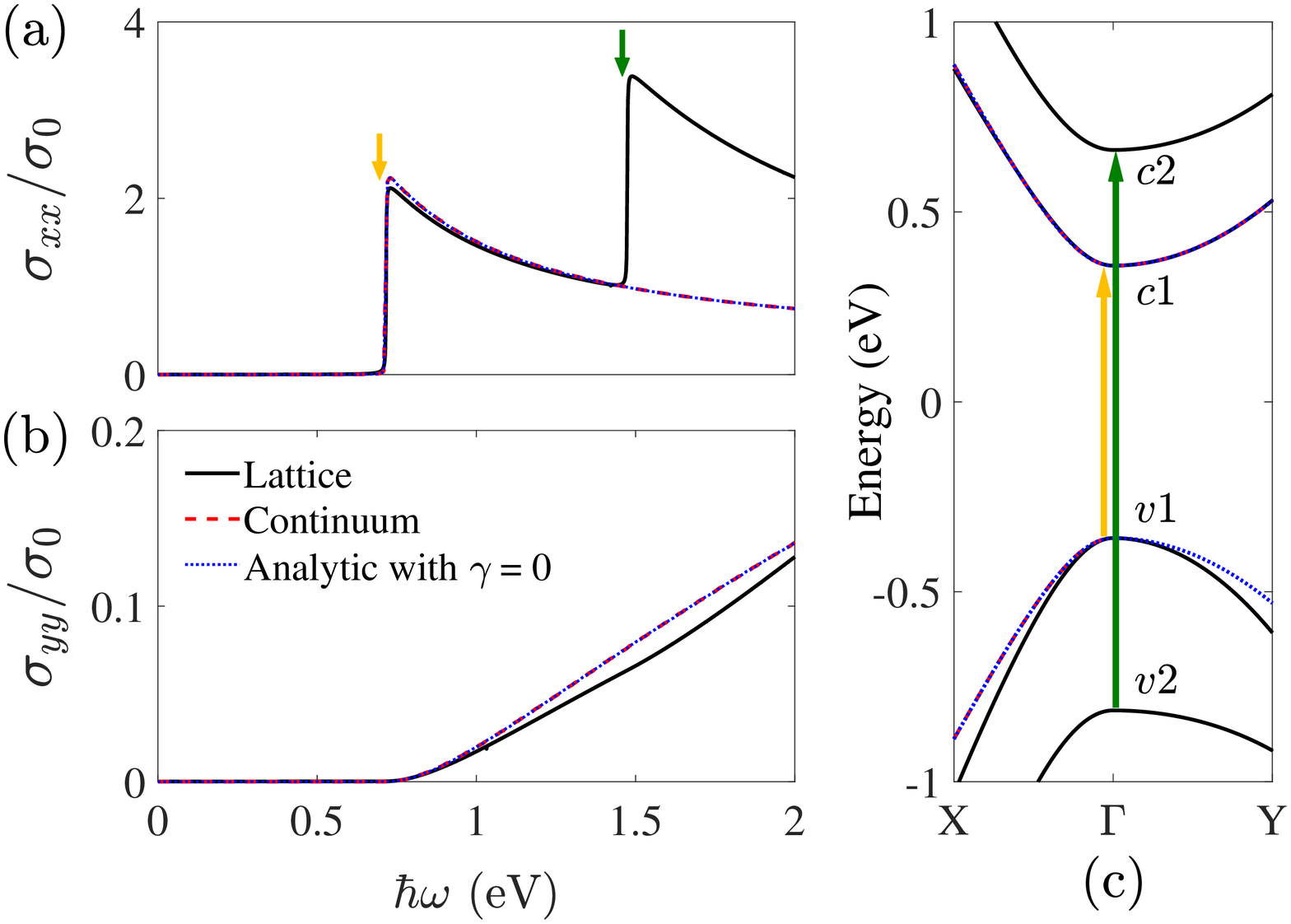}
\caption{
Optical conductivities (a) $\sigma_{xx}$ and (b) $\sigma_{yy}$ of tetralayer BP in the insulator phase with zero $E_{\rm ext}$ for the lattice model (black solid line), continuum model (red dashed line), and analytic result with $\gamma=0$ (blue dotted line). (c) The band structure of tetralayer BP in the insulator phase. Arrows indicate the interband transitions corresponding to the kink structures in $\sigma_{xx}$.
Here, $\sigma_0={e^2\over 4\hbar}$, and we adopt the following parameters for the calculation: $E_{\rm{ext}}=0$ V/$\rm \AA$ (pristine case), $\varepsilon_{\rm g}=0.717$ eV, $m=0.89m_{\rm e}$, $\gamma = 0$ and $v=6.2 \times 10^7$ cm/s.
}
\label{fig:optical_conductivity_normal_insulator}
\end{figure}
%%%%%%%%%%%%%%%%%%%%%%%%%%%%%%%%%%%%%%%%%%%%%%%%%%

Figure \ref{fig:optical_conductivity_normal_insulator} illustrates the calculated optical conductivities in the insulator phase for the lattice (black solid line) and corresponding continuum (red dashed line) models. The blue dotted line represents the analytic result with $\gamma=0$ obtained in the vicinity of $\hbar\omega=\varepsilon_{\rm g}$:
% \begin{subequations}
% \begin{eqnarray}
% \nonumber
% \sigma_{xx}(\omega)&\approx& g_{\rm s} \frac{e^2}{4\hbar}
% \sqrt{\frac{\varepsilon_0}{\varepsilon_{\rm g}}}
% \left[
% \frac{1}{2}-\frac{11}{16\varepsilon_{\rm g}}
% \left(\hbar \omega -\varepsilon_{\rm g} \right)
% \right]\\
% &&\times \Theta(\hbar\omega-\varepsilon_{\rm g}), \\
% \nonumber
% \sigma_{yy}(\omega)&\approx& g_{\rm s} \frac{e^2}{4\hbar}
% \left(\frac{\varepsilon_0}{\varepsilon_{\rm g}}\right)^{3/2}
% \left(\frac{\hbar \omega -\varepsilon_{\rm g}}{\varepsilon_0}\right)^2
% \\
% &&\times\Theta(\hbar\omega-\varepsilon_{\rm g}),
% \end{eqnarray}
% \end{subequations}
\begin{subequations}
\begin{eqnarray}
\!\!\!\!\sigma_{xx}(\omega)\!\!&\approx&\!\! \frac{g_{\rm s} e^2}{4\hbar}
\!\sqrt{\frac{\varepsilon_0}{\varepsilon_{\rm g}}}
\left[
\frac{1}{2}\!-\!\frac{11(\hbar \omega -\varepsilon_{\rm g})}{16\varepsilon_{\rm g}}
\right]
\!\Theta(\hbar\omega-\varepsilon_{\rm g}), \\
\!\!\!\!\sigma_{yy}(\omega)\!\! &\approx&\!\! \frac{g_{\rm s} e^2}{4\hbar}
\!\left(\frac{\varepsilon_0}{\varepsilon_{\rm g}}\right)^{3/2}
\left(\frac{\hbar \omega -\varepsilon_{\rm g}}{\varepsilon_0}\right)^2
\!\Theta(\hbar\omega-\varepsilon_{\rm g}),
\end{eqnarray}
\end{subequations}
where $\varepsilon_0 =  2m_y v^2$, $g_{\rm s}=2$ accounts for spin degeneracy, and $\Theta(x)$ is the step function, with $\Theta(x)=1$ for $x>0$ and 0 otherwise. %It is important to note that the continuum and analytic results agree well with the lattice results both qualitatively and quantitatively.

The energy gap with size $\varepsilon_{\rm g}$ leads to zero conductivity for frequencies $\hbar\omega<\varepsilon_{\rm g}$, owing to the absence of interband transitions. At the onset of interband transitions at $\hbar\omega=\varepsilon_{\rm g}$, $\sigma_{xx}$ exhibits a sudden jump and then decreases linearly, while $\sigma_{yy}$ increases quadratically with an increasing frequency $\omega$. %($\sigma_{yy}\sim\omega^2$).

It is also worth noting that $\sigma_{xx}$ exhibits two distinct kink structures at $\hbar\omega=0.72$ eV and $\hbar\omega=1.48$ eV, which are attributed to interband transitions between states near the $\Gamma$ point, as indicated by yellow ($v_1 \rightarrow c_1$) and green ($v_2 \rightarrow c_2$) arrows in Fig.~\ref{fig:optical_conductivity_normal_insulator}(c), whereas other interband transitions ($v_1 \rightarrow c_2$ and $v_2 \rightarrow c_1$) are forbidden \cite{low2014tunable, lin2016multilayer}. Along the zigzag direction, $\sigma_{yy}$ lacks such features because the interband transitions around the $\Gamma$ point are suppressed owing to the selection rule \cite{yuan2015polarization}.
%As mentioned, in the absence of a perpendicular field, there are only two strong interband transitions in $\sigma_{xx}$ indicated by yellow and green arrows ($v_1 \rightarrow c_1$, $v_2 \rightarrow c_2$), and other interband transitions indicated by blue and brown arrows ($v_2 \rightarrow c_1$, $v_1 \rightarrow c_2$) are forbidden \cite{low2014tunable, lin2016multilayer}, where $c_i$ ($v_i$) denotes conduction (valence) bands and subscript 1 describes the closer band to the Fermi level than the subscript 2.

%%%%%%%%%%%%%%%%%%%%%%%%%%%%%%%%%%%%%%%%%%%%%%%%%%
\begin{figure}[!h]
\includegraphics[width=1.0\linewidth]{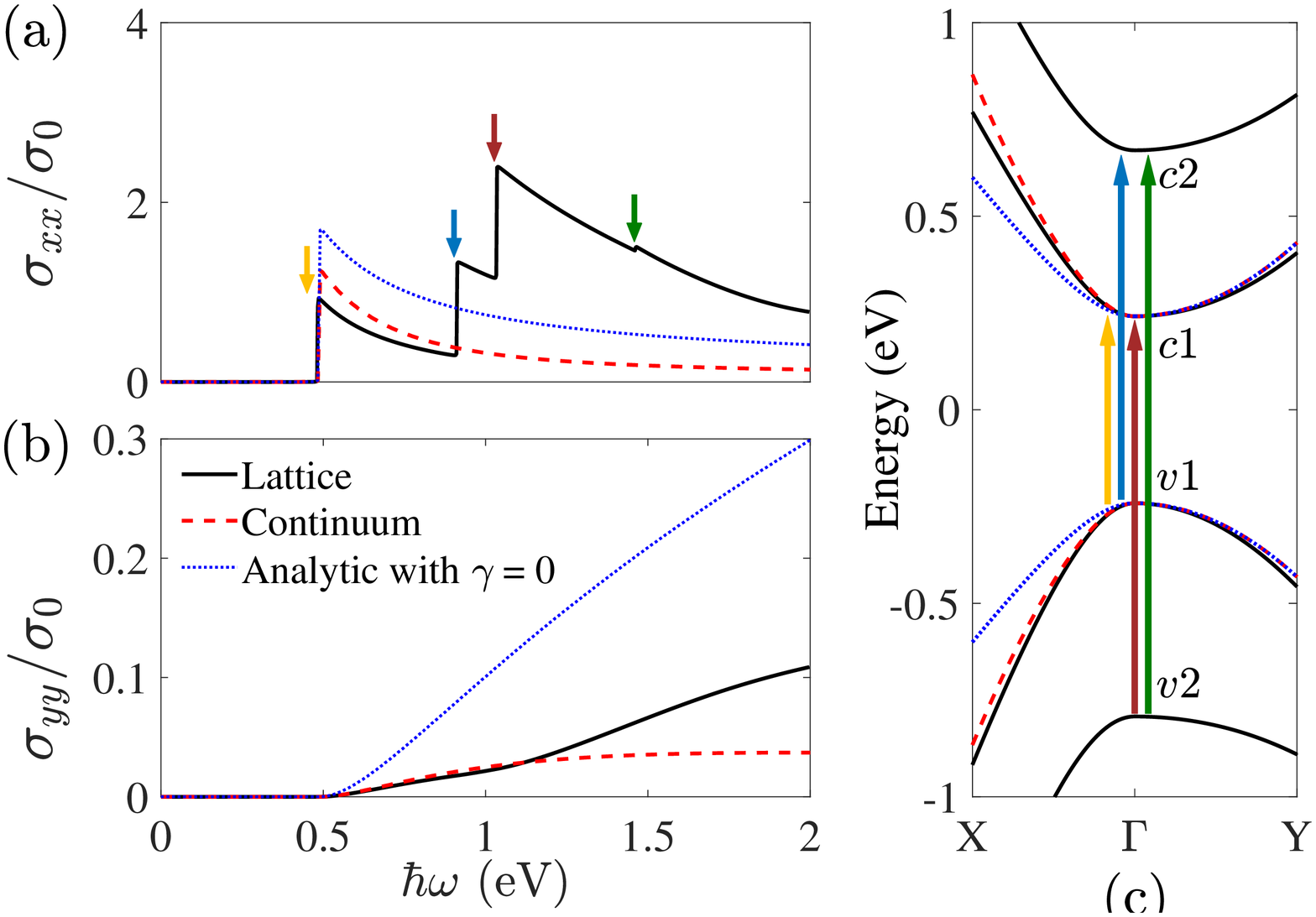}
\caption{
Optical conductivities (a) $\sigma_{xx}$ and (b) $\sigma_{yy}$ of tetralayer BP in the insulator phase with finite $E_{\rm ext}$ for the lattice model (black solid line), continuum model (red dashed line), and analytic result with $\gamma=0$ (blue dotted line). (c) The band structure of tetralayer BP in the insulator phase. Arrows indicate the interband transitions corresponding to the kink structures in $\sigma_{xx}$.
Here, $\sigma_0={e^2\over 4\hbar}$, and we adopt the following parameters for the calculation: $E_{\rm{ext}}=0.1$ V/\AA, $\varepsilon_{\rm g}= $ 0.483 eV, $m= 0.90$ $m_{\rm e}$, $\gamma= 2.6$, and $v=3.8 \times 10^7$ cm/s.
}
\label{fig:optical_conductivity_normal_insulator2}
\end{figure}
%%%%%%%%%%%%%%%%%%%%%%%%%%%%%%%%%%%%%%%%%%%%%%%%%%

However, a perpendicular electric field breaks the symmetry which is responsible for the two forbidden transitions indicated by blue ($v_1 \rightarrow c_2$) and brown ($v_2 \rightarrow c_1$) arrows in Fig.~\ref{fig:optical_conductivity_normal_insulator2}(c). Thus these forbidden interband transitions are now allowed in $\sigma_{xx}$ \cite{lin2016multilayer}.
Note that $\mathcal{M}_y$ is still preserved in the presence of a perpendicular electric field, and thus the optical conductivity of biased few-layer BP exhibits suppression in $\sigma_{yy}$ near $\hbar\omega=\varepsilon_{\rm g}$.
This result qualitatively agrees with recent experiments on the optical response of few-layer BP under a perpendicular electric field
\cite{doi:10.1021/acs.nanolett.6b03362, doi:10.1021/acs.nanolett.7b03050}.
The evolution of the optical peaks with the electric field and a detailed explanation of the selection rule are given in Appendix~\ref{app:evolution}.
%[{\bf Ahn: this paragraph needs to be supplemented with a figure or some other results. Any results presented above are not really related to the main point of this paragraph.}]
%[{\bf Min: We can add explanation on the evolution of the peaks with $E_{\rm ext}$ between the insulator phase and the semi-Dirac point. If we can find the locations of the peak analytically, we can include them. What is the symmetry responsible for the forbidden transitions? We need to explain this more clearly.}]

%{\bf [Min: I think ``As mentioned in .. (see Appendix A)" part can be moved to Appendix.]}
%we can conclude that the reflection symmetry about the $y=0$ plane suppresses peaks in $\sigma_{yy}$.
%[{\bf Jang: What does it mean?}]
%arising from the diverging joint density of states near the van Hove singularity.
%Thus, in the vicinity of the $\Gamma$ point, the matrix element along the zigzag direction ($M^{ss'}_y(\bm k)$) is small compared to the one along the armchair direction ($M^{ss'}_x(\bm k)$). Since $M_y$ is preserved in the presence of a perpendicular electric field, we can conclude that biased few-layer BP also exhibits strong optical anisotropy.

As forbidden interband transitions appear, the oscillator strength of the optical transitions, indicated by yellow ($v_1 \rightarrow c_1$) and green arrows ($v_2 \rightarrow c_2$), is reduced [Fig.~\ref{fig:optical_conductivity_normal_insulator2}(a)]. However, the continuum model and analytic result cannot capture this reduced oscillator strength, because they are effective two-band models at low energies, only including transitions between the two bands.

%We confirm that our result qualitatively agrees with recent experimental studies of the optical response of few-layer BP with a perpendicular electric field \cite{doi:10.1021/acs.nanolett.6b03362, doi:10.1021/acs.nanolett.7b03050}.

%%%%%%%%%%%%%%%%%%%%%%%%%%%%%%%%%%%%%%%%%%%%%%%%%%

\subsubsection{Semi-Dirac point}
\label{subsec:optical_conductivity_semi_Dirac}
The zero-temperature optical conductivity with $\gamma=0$ at the semi-Dirac point ($\varepsilon_{\rm g}=0$) is given by
%\begin{subequations}
%\label{semidirac_analytic}
%\begin{align}
%\sigma_{xx}(\omega)&=g_{\rm s}{e^2\over 4\hbar}\
%A_{xx} \left(\frac{\omega}{\omega_0}\right)^{-{1\over 2}}\Theta(\omega-2|\omega_{\mu}|),\\
%\sigma_{yy}(\omega)&=g_{\rm s}{e^2\over 4\hbar}\
%A_{yy} \left(\frac{\omega}{\omega_0}\right)^{{1\over 2}}\Theta(\omega-2|\omega_{\mu}|),
%\end{align}
%\end{subequations}
 \begin{subequations}
 \label{semidirac_analytic}
 \begin{flalign}
 &\sigma_{xx}(\omega)=g_{\rm s} {e^2\over 4\hbar}\\
 &\times \left[A_{xx}\left(\frac{|\mu|}{\varepsilon_0}\right)^{\frac{1}{2}} \delta(\hbar \omega) + B_{xx} \left(\frac{\omega}{\omega_0}\right)^{-{1\over 2}}\Theta(\omega-2|\omega_{\mu}|)\right], \nonumber \\
 &\sigma_{yy}(\omega)= g_{\rm s}{e^2\over 4\hbar}\\
 &\times \left[A_{yy}\left(\frac{|\mu|}{\varepsilon_0}\right)^{\frac{3}{2}} \delta(\hbar \omega) + B_{yy} \left(\frac{\omega}{\omega_0}\right)^{{1\over 2}}\Theta(\omega-2|\omega_{\mu}|)\right], \nonumber
\end{flalign}
\end{subequations}
% \begin{subequations}
% \label{semidirac_analytic}
% \begin{eqnarray}
% {\sigma_{xx}(\omega)\over \sigma_0}\!\!&=&\!\!\! A_{xx}\left(\frac{|\mu|}{\varepsilon_0}\right)^{\frac{1}{2}} \delta(\hbar \omega) + B_{xx} \left(\frac{\omega}{\omega_0}\right)^{-{1\over 2}} \!\!\! \Theta(\omega-2|\omega_{\mu}|), \\
% {\sigma_{yy}(\omega)\over \sigma_0}\!\!&=&\!\!\! A_{yy}\left(\frac{|\mu|}{\varepsilon_0}\right)^{\frac{3}{2}} \delta(\hbar \omega) + B_{yy} \left(\frac{\omega}{\omega_0}\right)^{{1\over 2}}\Theta(\omega-2|\omega_{\mu}|),
% \end{eqnarray}
% \end{subequations}
%where $\mu$ is the chemical potential, $\omega_\mu=\mu/\hbar$, $A_{xx}=\dfrac{{\mathrm B}(\frac{1}{4},\frac{1}{4})}{3 \pi}$, $A_{yy}=\dfrac{8}{5 {\mathrm B}(\frac{1}{4},\frac{1}{4})}$. Here, $\mathrm{B}(n,m)$ is the beta function. (See Appendix \ref{sec:analytic_expressions} for the detailed derivations.)
where $\mu$ is the chemical potential, $\omega_\mu=\mu/\hbar$, $A_{xx} = \frac{2}{3\sqrt{\pi}}\frac{\Gamma{(1/4)}}{\Gamma{(3/4)}}$, $A_{yy} = \frac{48}{5\sqrt{\pi}}\frac{\Gamma{(3/4)}}{\Gamma{(1/4)}}$, $B_{xx}=\frac{1}{6\sqrt{2\pi}} \frac{\Gamma(1/4)}{\Gamma(3/4)}$, and $B_{yy}=\frac{4\sqrt{2}}{5\sqrt{\pi}} \frac{\Gamma(3/4)}{\Gamma(1/4)}$. Here, $\Gamma(z)=\int_0^{\infty}t^{z-1}e^{-t} dt$ is the gamma function. (See Appendix \ref{sec:analytic_expressions} for detailed derivations.)
%{\bf [Min: Since we mention the intraband transition below, we had better recover the original expressions including the intraband contributions.]}

The first term represents intraband transitions, giving rise to the Drude peak at low frequencies. The second term represents interband transitions, which are forbidden at $\omega<2|\omega_\mu|$ owing to Pauli blocking. Because a finite $\mu$ simply leads to Pauli blocking for interband transitions and the Drude peak for intraband transitions, from now on we only consider the undoped case with $\mu=0$.
In the undoped case, the optical conductivity at low frequencies scales as a power law, with $\sigma_{xx}(\omega)\propto\omega^{-{1\over 2}}$ (armchair direction) and $\sigma_{yy}(\omega)\propto\omega^{{1\over 2}}$ (zigzag direction), which is consistent with previous studies \cite{cho2016novel, isobe2016emergent}.

%%%%%%%%%%%%%%%%%%%%%%%%%%%%%%%%%%%%%%%%%%%%%%%%%%
\begin{figure}[t]
\includegraphics[width=1.0\linewidth]{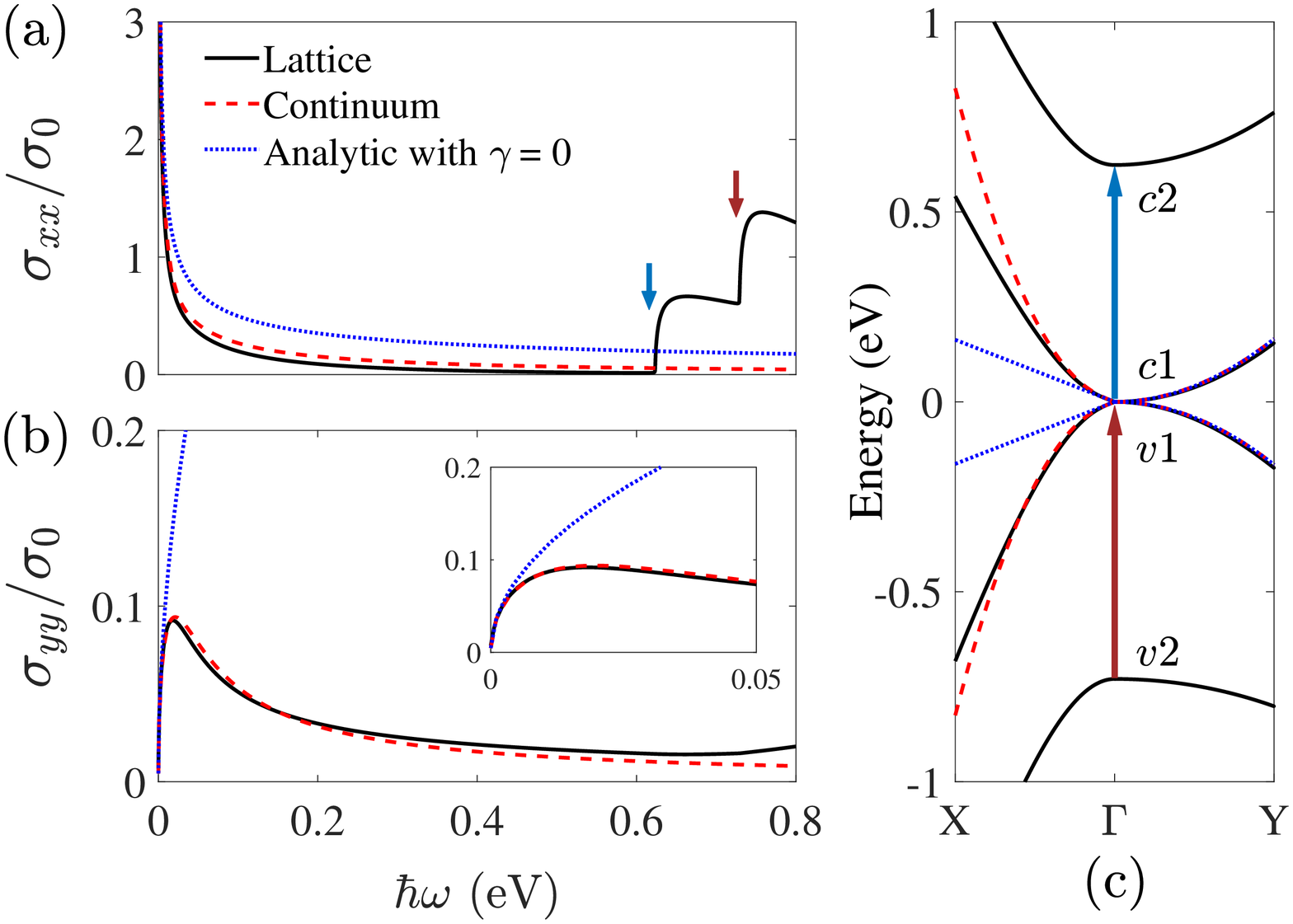}
\caption
{
Optical conductivities (a) $\sigma_{xx}$ and (b) $\sigma_{yy}$ of tetralayer BP at the semi-Dirac point for the lattice model (black solid line), continuum model (red dashed line), and analytic result with $\gamma=0$ (blue dotted line). (c) The band structure of tetralayer BP at the semi-Dirac point. Arrows indicate the interband transitions corresponding to the kink structures in $\sigma_{xx}$. Here, $\sigma_0={e^2\over 4\hbar}$, and we adopt the following parameters for the calculation: $E_{\rm{ext}}=0.2141$ V/\AA, $\varepsilon_{\rm g}=0$ eV, $m=0.92m_{\rm e}$, $\gamma = 4.9$, and $v=1.2 \times 10^7$ cm/s.
}
\label{fig:optical_conductivity_semi_Dirac}
\end{figure}
%%%%%%%%%%%%%%%%%%%%%%%%%%%%%%%%%%%%%%%%%%%%%%%%%%

Figure \ref{fig:optical_conductivity_semi_Dirac} illustrates the calculated optical conductivities at the semi-Dirac point for the lattice and continuum models, along with the analytic result with $\gamma=0$ [Eq.~(\ref{semidirac_analytic})]. As in the case for the insulator phase, the three results are in good agreement at low frequencies.
% But at around $\hbar\omega_c=0.03$ eV ($\approx 7.3$ THz), the analytic result starts deviating from both the continuum and the lattice results, which is particularly striking for $\sigma_{yy}$: the analytic result continually increases beyond $\hbar\omega_c$, while the other two monotonically decreases at $\omega>\omega_c$.
% Such deviations can be explained with the band structures presented in Fig.~ \ref{fig:optical_conductivity_semi_Dirac}(c).  At low frequencies, the lattice Hamiltonian is well approximated by the corresponding low-energy model in Eq. ~(\ref{eq:continuum_model_2}).
% [{\bf Ahn: It needs an explanation on how $\gamma$ affects the energy band and thus the optical conductivity}]
However, as the frequency increases the analytic result begins to deviate from both the continuum and lattice results, which is particularly striking for $\sigma_{yy}$: here, the analytic result continues to increase with the $\omega^{1\over 2}$ dependence, while the other two results monotonically decrease with the frequency. Such a deviation can be explained by the effect of the parabolic term $\gamma {\hbar^2 k_x^2\over 2 m}$ in Eq.~(\ref{eq:continuum_model_2}). Because there are both linear ($\hbar v k_x$) and parabolic ($\gamma {\hbar^2 k_x^2\over 2 m}$) terms in $k_x$, there exists a crossover energy $\hbar\omega_{\rm cr}=\hbar v k_x=\gamma {\hbar^2 k_x^2\over 2 m}$, which is given by $\hbar\omega_{\rm cr}={2 m v^2 \over \gamma}$. For $\omega\ll \omega_{\rm cr}$, the linear term is dominant, and the optical conductivity exhibits $\sigma_{yy}\sim \omega^{1/2}$, as obtained in Eq.~(\ref{semidirac_analytic}b) by neglecting the parabolic term ($\gamma=0$). However, for $\omega\gg\omega_{\rm cr}$ the parabolic term is dominant, and $\sigma_{yy}\sim \omega^{-1}$.
%(see Appendix \ref{app:evolution}).
The effect of $\gamma$ becomes significant as the phase changes from the insulator to the Dirac semimetal phase, as shown in Fig.~\ref{fig:parameter_evolution}(b).
%{\bf [Min: Check that $\omega_{\rm cr}$ becomes significant or remains almost the same.]}

% \textbf{(Explain based on the $\gamma$ term, ... Energy band dispersion of biased few-layer BP at the bandgap closing point is often approximated as the \textit{semi-Dirac} Hamiltonian, which corresponds to $\widetilde{\varepsilon}_g=0$ and $\gamma=0$ case in Eq. ~(\ref{eq:continuum_model_2_dimensionless}) \cite{doh2017dirac,baik2015emergence, de2017anisotropic}. In a low energy regime, the semi-Dirac Hamiltonian describes few-layer BP quite well since the optical conductivity show the monotonic decreasing and increasing behavior of $\sigma_{xx}$ and $\sigma_{yy}$ with the power $\omega^{\mp \frac{1}{2}}$ as obtained from the Reference \cite{cho2016novel, isobe2016emergent}.}

\subsubsection{Dirac semimetal phase}
\label{subsec:optical_conductivity_semi_metallic}

The zero-temperature optical conductivity with $\gamma=0$ for the Dirac semimetal phase ($\varepsilon_{\rm g}<0$) at low frequencies with $\mu=0$ is given by
\begin{subequations}
\label{eq:semimetalic_analytic}
\begin{eqnarray}
\sigma_{xx}(\omega)&\approx& 2g_{\rm s}{e^2\over 16\hbar}{v_x\over v_y}+g_{\rm s}{e^2\over 4\hbar}C_{xx}\left(\frac{\omega}{\omega_0}\right)^{2},\\
\sigma_{yy}(\omega)&\approx& 2g_{\rm s}{e^2\over 16\hbar}{v_y\over v_x}+g_{\rm s}{e^2\over 4\hbar}C_{yy}\left(\frac{\omega}{\omega_0}\right)^{2},
\end{eqnarray}
\end{subequations}
where $v_x=v$, $v_y = \frac{\hbar k_{\rm D}}{m_y} =\sqrt{\frac{|\varepsilon_{\rm g}|}{m_y}}$ with $k_{\rm D}= \sqrt{\frac{m_y |\varepsilon_{\rm g}|}{\hbar^2}}$, $C_{xx}=\frac{9}{64\sqrt{2}} \left(\frac{\varepsilon_0}{|\varepsilon_{\rm g}|}\right)^{5/2}$, and $C_{yy}=-\frac{1}{32\sqrt{2}} \left(\frac{\varepsilon_0}{|\varepsilon_{\rm g}|}\right)^{3/2}$.

As $\varepsilon_{\rm g}$ decreases below zero, the semi-Dirac point located at the $\Gamma$ point is split into two Dirac nodes at $\bm{k} = ({0, \pm k_{\rm D}})$ (see Fig.~\ref{fig:electronic_structure}). Thus, the optical conductivities in the zero-frequency limit can be interpreted as the sum of the optical conductivities from the two independent 2D Dirac nodes (such as graphene) with the anisotropic in-plane velocities $v_x$ and $v_y$. (See Appendix \ref{sec:analytic_expressions} for detailed derivations.)
Note that these velocities vary with the electric field, as shown in Fig.~\ref{fig:parameter_evolution}.
%[{\bf Min: Change the notations in Eq.~(\ref{eq:semimetalic_analytic}) following Eq.~(\ref{eq:continuum_model_2}). For the evolution of $v_y$, we can either include it in Fig.~\ref{fig:parameter_evolution} or in the Appendix \ref{app:evolution}.}]
%%%%%%%%%%%%%%%%%%%%%%%%%%%%%%%%%%%%%%%%%%%%%%%%%%

\begin{figure}[!htb]
\includegraphics[width=1\linewidth]{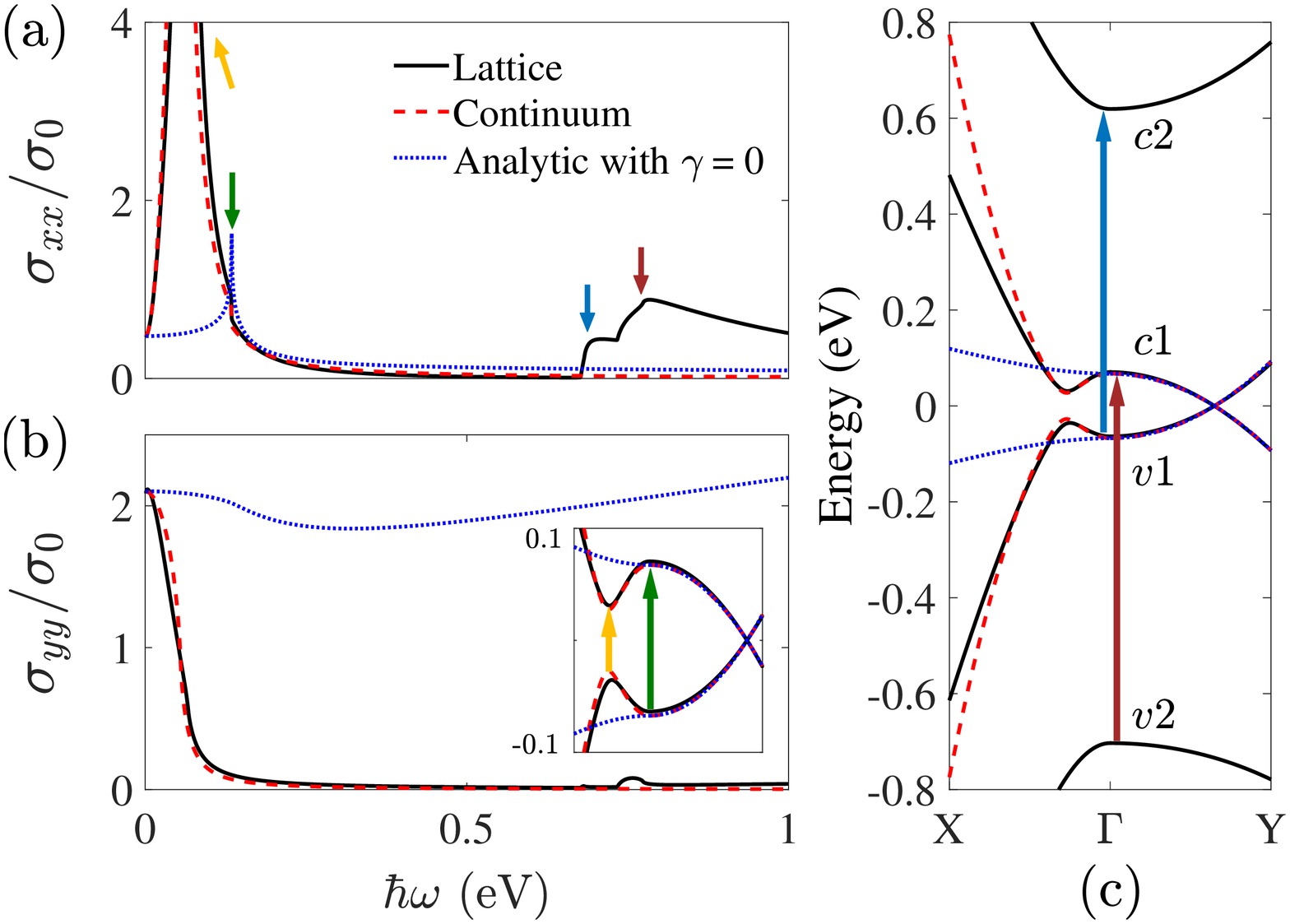}
\caption
{
Optical conductivities (a) $\sigma_{xx}$ and (b) $\sigma_{yy}$ of tetralayer BP in the Dirac semimetal phase for the lattice model (black solid line), continuum model (red dashed line), and analytic result with $\gamma=0$ (blue dotted line). (c) The band structure of tetralayer BP in the Dirac semimetal phase. Arrows indicate the interband transitions corresponding to the kink structures in $\sigma_{xx}$. Here, $\sigma_0={e^2\over 4\hbar}$ and we adopt the following parameters for the calculation: $E_{\rm{ext}}=0.4$ V/\AA, $\varepsilon_{\rm g}=-0.134$ eV, $m = 0.95m_{\rm e}$, $\gamma = 5.2$ and $v=7.5 \times 10^6$ cm/s.
%{\bf [Min: If (c) looks crowded, we can move the inset in (c) to (b).]}
}
\label{fig:optical_conductivity_semi_metallic}
\end{figure}
%%%%%%%%%%%%%%%%%%%%%%%%%%%%%%%%%%%%%%%%%%%%%%%%%%
%Figure \ref{fig:QAH_optical_conductivity} shows calculated optical conductivities for the $J=1$ and $J=2$ lattice and continuum models in the 3D QAH phase. If $\gamma=0$, the energy gap with a size of $2|\alpha|$ for both NI and 3D QAH phases leads to zero conductivity for frequencies $\hbar\omega<2|\alpha|$ due to the optical gap. Because of the non-zero $\gamma$, a Mexican hat structure appears in the 3D QAH phase (but not in the NI phase) if $\alpha<\alpha_{\rm c}=-{\varepsilon_0^2 \over 2\gamma k_0^2}$ for $J=1$, and if $\alpha<0$ for $J=2$ exhibiting a shifted interband peak with respect to the $\gamma=0$ result \cite{Supplemental}. For the $J=1$ lattice model in the 3D QAH phase, an additional kink structure appears at $\hbar\omega=2|m_z-t_z+2m_0|$ due to the interband transitions at local minima $(k_x,k_y,k_z)=(\pm {\pi\over a},0,0)$, $(0,\pm {\pi\over a},0)$, as shown in Fig.~\ref{fig:QAH_optical_conductivity}(a).

Figure \ref{fig:optical_conductivity_semi_metallic} illustrates the optical conductivities calculated in the Dirac semimetal phase for the lattice and continuum models, along with the analytic result obtained by assuming that $\gamma=0$  [Eq.~(\ref{eq:semimetalic_analytic})].
If $\gamma=0$, then an optical peak in $\sigma_{xx}$ occurs at $\hbar\omega=|\varepsilon_{\rm g}|$, owing to the interband transition at the $\Gamma$ point. However, as shown in Fig.~\ref{fig:parameter_evolution}, $\gamma$ increases as the phase changes from the insulator to the Dirac semimetal phase. Thus, it is expected that the analytic result with $\gamma=0$ will exhibit a deviation from those of the lattice and continuum models. For non-zero $\gamma$, the band structure is modified and a shifted interband transition occurs away from the $\Gamma$ point, exhibiting an optical peak at $\omega=\omega_{\rm cr}\sqrt{{2|\varepsilon_{\rm g}|\over \hbar\omega_{\rm cr}}-1}$ (which is typically less than the order of 100 meV for few-layer BP) if $\hbar\omega_{\rm cr}<2|\varepsilon_{\rm g}|$ (or equivalently $\gamma>{m v^2\over |\varepsilon_{\rm g}|}$). For tetralayer BP, the optical peak occurs around $\hbar \omega = 0.067$ eV, as indicated by the yellow arrow in Fig.~\ref{fig:optical_conductivity_semi_metallic}. The analytic result with $\gamma=0$ cannot capture this peak, because the band structure with $\gamma=0$ does not show van Hove singularities other than the $\Gamma$ point, exhibiting only one peak around $\hbar \omega = 0.134$ eV, as indicated by the green arrow. Kink structures in $\sigma_{xx}$ at higher frequencies are indicated by blue and brown arrows.

It is worth noting that the low-frequency optical conductivity for non-zero $\gamma$ has the same form in Eq.~(\ref{eq:semimetalic_analytic}) as obtained for $\gamma=0$, but with different coefficients $C_{xx}$ and $C_{yy}$. We found that both $C_{xx}$ and $C_{yy}$ are enhanced by increasing $\gamma$, whereas $v_x$ and $v_y$ do not change with $\gamma$, giving the same optical conductivity in the zero-frequency limit irrespective of $\gamma$.

The optical conductivity along the zigzag direction $\sigma_{yy}$ does not exhibit any peaks in $\sigma_{xx}$, because the vanishing matrix elements forbid such transitions. As the frequency increases, a discrepancy in $\sigma_{yy}$ between the analytic result and the results of the two other models becomes significant. Both the lattice and continuum model results decrease, whereas the analytic result increases, owing to the absence of the parabolic term $\gamma {\hbar^2 k_x^2\over 2 m}$.

\subsection{Finite temperature effect}
%[{\bf Ahn: If the finite temperature results are only meaningful for the semi-metallic phase, then I think we can just put them in a subsection of the previous section instead of creating a new section.}]
So far, we have focused on the zero temperature case. In this section, we analyze the effect of a finite temperature on the optical conductivity at low frequencies.
%In case of the semi-Dirac point and the Dirac phase, the thermal energy is comparable to the regime where low-energy optical transitions occur. Thus, it is worthwhile to mention the contribution of finite temperature to the optical conductivity.
If conduction and valence bands are symmetric in a two-band model, then the optical conductivity at finite temperature is reduced to a compact form as follows (see Appendix \ref{sec:analytic_expressions} for details):
\begin{eqnarray}
\sigma_{ii}(\omega,T,\mu) = A(\omega,T,\mu) \sigma_{ii}(\omega,T=0,\mu=0),
\end{eqnarray}
where
% \begin{eqnarray}
% A(\omega,T,\mu)=\frac{\sinh{(\hbar\omega/2k_{\rm B}T)}}{\cosh{(\hbar\omega/2k_{\rm B}T)}+\cosh{(\mu/k_{\rm B}T)}}.
% \end{eqnarray}
\begin{align}
A(\omega,T,\mu)=\frac{\sinh{\left({\hbar\omega\over 2k_{\rm B}T}\right)}}{\cosh{\left({\hbar\omega \over 2k_{\rm B}T}\right)}+\cosh{\left({\mu\over k_{\rm B}T}\right)}}.
\end{align}
Note that $A(\omega,T=0,\mu)=\Theta(\hbar \omega - 2|\mu|)$ for $T=0$, reproducing the zero-temperature result with Pauli blocking. Furthermore, note that for $\mu=0$, $A(\omega,T,\mu=0)=\tanh{\left({\hbar\omega\over 4k_{\rm B}T}\right)}$.

%In the case of the insulator phase, typically the thermal energy is negligible compared to the band gap. For the Dirac phase, the thermal energy is comparable to the regime where low-energy optical transitions occur. However, we observe that there is no significant modification of the optical conductivity.
%[{\bf Min: Not sure the argument for the Dirac phase is true. The finite temperature will also modify the result by the same factor, as shown above, though the observable range could be different. I think in the Dirac semimetal phase, we may observe the similar power-law change while in the insulator phase, the finite temperature effect is typically negligible. Here, we present the result of the semi-Dirac point for demonstration.}]

%%%%%%%%%%%%%%%%%%%%%%%%%%%%%%%%%%%%%%%%%%%%%%%%%%
\begin{figure}[t]
\includegraphics[width=1.0\linewidth]{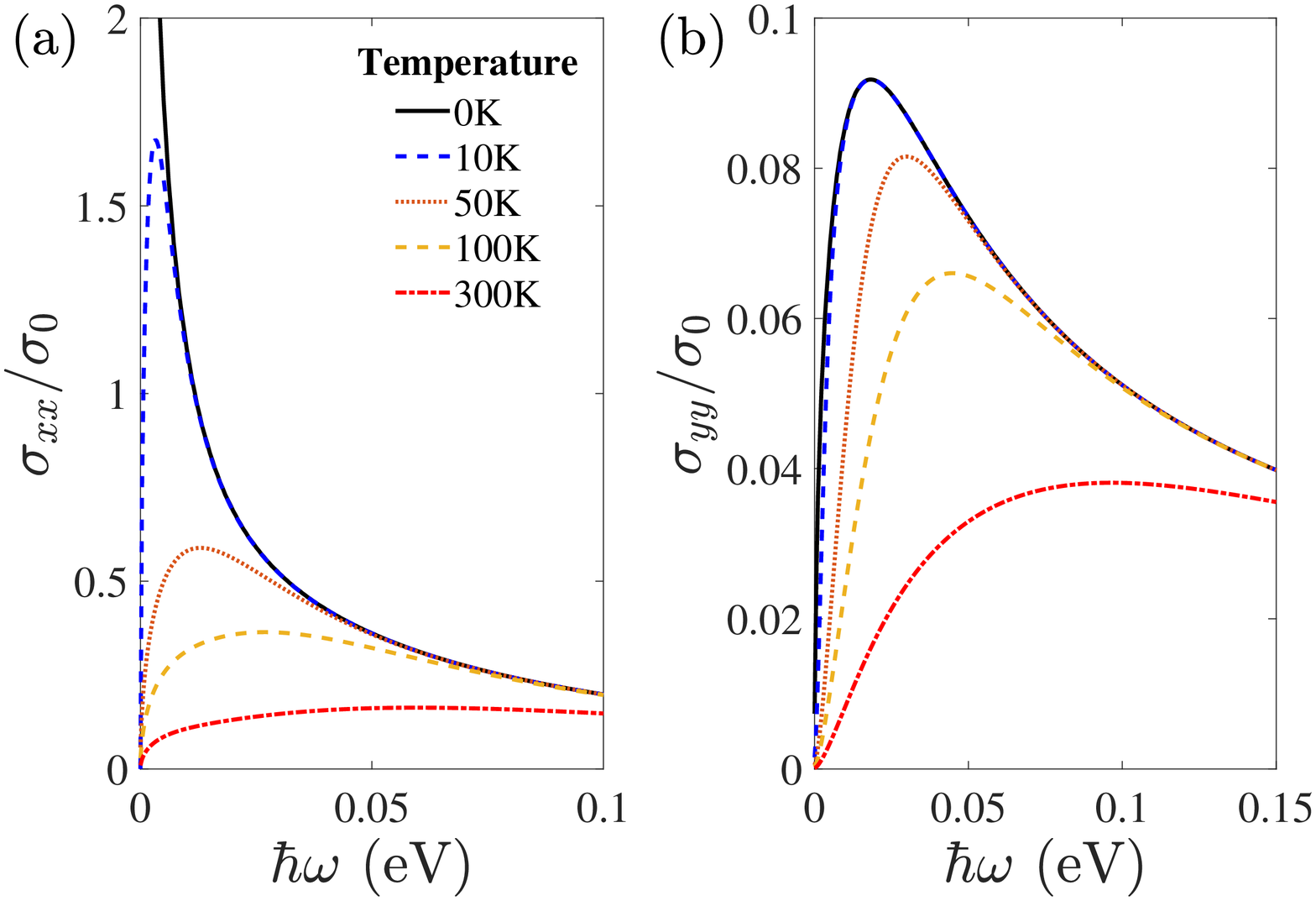}
\caption{
Optical conductivities (a) $\sigma_{xx}$ and (b) $\sigma_{yy}$ of tetralayer BP obtained using the lattice model at the semi-Dirac point for various temperatures with $\mu=0$.
Here, we adopt the same parameters as in Fig.~\ref{fig:optical_conductivity_semi_Dirac} for the calculation.
%Here, $\sigma_0={e^2\over 4\hbar}$ and we use the following parameters for the calculation: $\cdots$
}
\label{fig:optical_conductivity_semi_Dirac_temp}
\end{figure}
%%%%%%%%%%%%%%%%%%%%%%%%%%%%%%%%%%%%%%%%%%%%%%%%%%
%%%%%%%%%%%%%%%%%%%%%%%%%%%%%%%%%%%%%%%%%%%%%%%%%%
\begin{figure}[!b]
\includegraphics[width=0.9\linewidth]{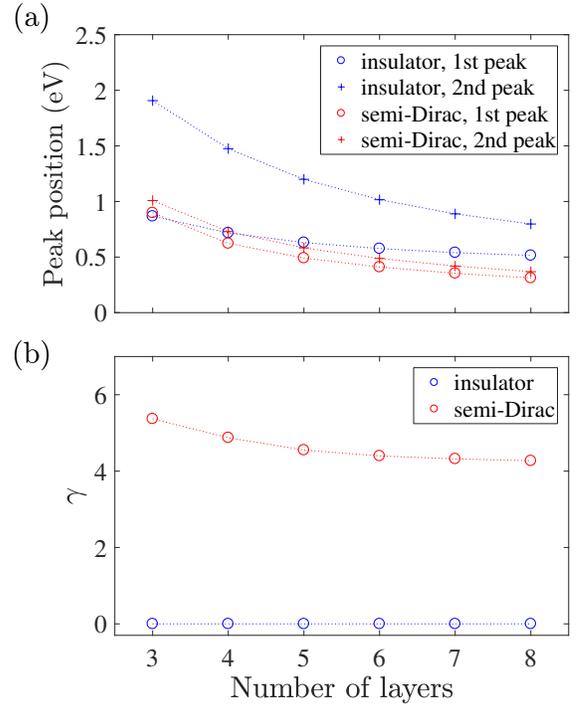}
\caption{(a) First and second optical peaks as a function of the number of layers in the insulator phase with $E_{\rm ext}=0$ and the semi-Dirac point.  (b) Evolution of the parameter $\gamma$ in the insulator phase with $E_{\rm ext}=0$ and the semi-Dirac point as a function of the number of layers.
}
\label{fig:layer_evol}
\end{figure}
%%%%%%%%%%%%%%%%%%%%%%%%%%%%%%%%%%%%%%%%%%%%%%%%%%

Figure~\ref{fig:optical_conductivity_semi_Dirac_temp} illustrates the optical conductivities for the semi-Dirac point calculated at finite temperature with $\mu=0$. At zero temperature, the low-frequency power-law of the optical conductivities at the semi-Dirac point is given by $\sigma_{xx}\sim \omega^{1/2}$ and $\sigma_{yy}\sim \omega^{-1/2}$. At finite temperature, the temperature factor $A(\omega,T,\mu=0)=\tanh{\left({\hbar\omega\over 4k_{\rm B}T}\right)}$ is multiplied, and so the power-law is modified to $\sigma_{xx}\sim \dfrac{\omega^{3/2}}{T}$ and $\sigma_{yy}\sim \dfrac{\omega^{1/2}}{T}$ for ${k_{\rm B}T}\gg \hbar\omega$. Here, we used $\tanh{(x)}\approx x$ for small $x$.
Similarly, in the insulator and Dirac semimetal phases, the optical conductivity at finite temperature exhibits a modified power-law at low frequencies.
%{\bf [Min: Check this.]}

%%%%%%%%%%%%%%%%%%%%%%%%%%%%%%%%%%%%%%%%%%%%%%%%%%%%%%%%%%%%%%%%%%%%%%%%%%%%%%%%%%%%%%%%%%%%%%%%%%%%
\section{Discussion and Summary}
\label{sec:discussion}

The present calculations are performed for tetralayer BP. As the number of layers increases, the electronic structure of few-layer BP and the corresponding optical conductivities change. Figure \ref{fig:layer_evol}(a) illustrates the evolution of the first and second optical peaks in the insulator phase with $E_{\rm ext}=0$ and at the semi-Dirac phase as the number of layers increases, showing that the peak positions decrease with the number of layers. We also demonstrate how the parameter $\gamma$ evolves as the number of layers increases in Fig.~(\ref{fig:layer_evol})(b). For the insulator phase with $E_{\rm ext}=0$, $\gamma$ remains around 0 regardless of the number of layers, whereas at the semi-Dirac point $\gamma$ decreases from 5.4 for three layers to 4.3 for eight layers of BP.

%The present calculations use model parameters obtained from the tetralayer BP.
%%{\bf [Min: Since we discuss the effect of the number of layers below, we can either focus on the insulator phase or the semi-Dirac point (or both), showing the first/second peaks and the evolution of $\gamma$. Currently, the cross-over energy is presented as a function of $\varepsilon_{\rm g}$ for different layers, and it seems comparison between different layer numbers with the same gap size is not so meaningful.]}
%As the number of layers increase, the gap decreases and the higher bands also approach the Fermi energy and thus the position of the optical peak would be shifted. Figure~(\ref{fig:layer_evol})(a) shows how the optical peak change with the number of layers. We present the two lowest optical peaks in the each insulator phase with zero external electric field and semi-Dirac point. We find a red shift of all four optical peaks as the number of layers increase. Along with the effect of the layers on the optical peaks, the evolution of the $\gamma$ with the number of layers is displayed in Fig~(\ref{fig:layer_evol})(b). For the insulator phase with zero external electric field, the $\gamma$ remains nearly 1 regardless of the number of layers, whereas at the semi-Dirac point, the $\gamma$ decreases from 4.6 for 3 layers of BP to 3.3 for 8 layers of BP. Since there is no abrupt change in the model parameters, our results for optical conductivity would be reliable for few-layer BP with an arbitrary numbers of layers.

%%%%%%%%%%%%%%%%%%%%%%%%%%%%%%%%%%%%%%%%%%%%%%%%%%
\begin{figure}[h]
\includegraphics[width=1.0\linewidth]{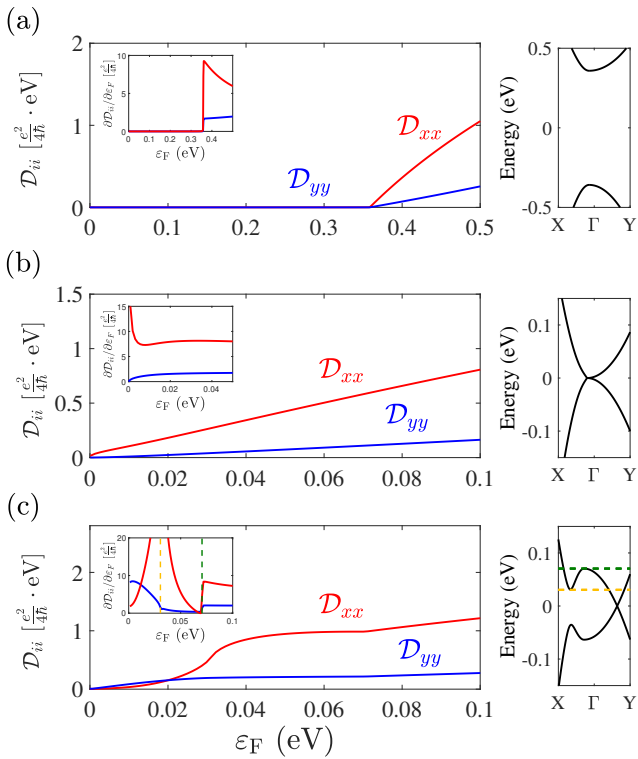}
\caption{
Drude weight as a function of Fermi energy for the lattice model in the (a) insulator phase, (b) semi-Dirac point and (c) Dirac semimetal phase. The right panels show the corresponding energy dispersions. The Drude weight is defined to be $\sigma_{ii} = \mathcal{D}_{ii} \delta(\hbar \omega)$. The insets show the derivatives of the Drude weight.
}
\label{fig:optical_conductivity_intraband}
\end{figure}
%%%%%%%%%%%%%%%%%%%%%%%%%%%%%%%%%%%%%%%%%%%%%%%%%%

As the Fermi energy moves away from zero, the intraband contribution to the optical conductivity arises at low frequencies. Figure \ref{fig:optical_conductivity_intraband} shows the Drude weight for each phase. As the Fermi energy increases, the Drude weight increases and exhibits kink structures at the van Hove singularities. These features can be observed more clearly in the derivatives of the Drude weight with respect to the Fermi energy, as shown in the insets of Fig.~\ref{fig:optical_conductivity_intraband}(c). Note that in the Dirac semimetal phase, the derivative of the Drude weight exhibits discontinuities at the van Hove singularities, indicated by yellow and green arrows.

In summary, we have studied the optical conductivity of a biased few-layer BP in each phase and at the corresponding transition points. In particular, we focused on the low-energy characteristic frequency dependence, which can be utilized as an experimental fingerprint. We analytically obtained the scaling law for the optical conductivity at low frequencies, and verified this using the corresponding full lattice model for few-layer BP. Beyond the low frequency regime, the analytic result exhibits a deviation from the result calculated based on the lattice model.
%We used the continuum model to calculate the optical conductivity and confirm that the continuum model can capture all the essential physics.
We systematically analyzed the role of the parabolic term $\gamma \frac{\hbar^2 k_x^2}{2m} \sigma_z$ in the optical conductivity. The parameter $\gamma$, which characterizes the contribution of the parabolic term, varies with the external electric field strength, and becomes significant as the phase changes from the insulator to the Dirac semimetal phase. At the semi-Dirac point, it was revealed that this parabolic term gives rise to a crossover frequency beyond which the low frequency scaling law ($\sigma_{xx}\sim \omega^{-1/2}$, $\sigma_{yy}\sim \omega^{1/2}$) is no longer valid. In the Dirac semimetal phase, the dominant low-frequency interband transition is shifted, owing to the band distortion associated with non-zero $\gamma$.
%and we obtained the evolution by using a self-consistent Hartree method. In addition, we investigated the finite temperature effect on the optical conductivity, which may change the low frequency power-law depending on the energy scale.

\acknowledgments
J.J. and H.M. were supported by the NRF grant funded by the Korea government (MSIT) (No. 2018R1A2B6007837) and Creative-Pioneering Researchers Program through Seoul National University. S.A. was supported by IBS-R009-D1 (G1, G2, G3, Y1).
%%%%%%%%%%%%%%%%%%%%%%%%%%%%%%%%%%%%%%%%%%%%%%%%%%%%%%%%%%%%%%%%%%%%%%%%%%%%%%%%%%%%%%%%%%%%%%%%%%%%

\begin{widetext}
\appendix
\section{Analytic expressions of optical conductivity for each phase}
\label{sec:analytic_expressions}
In the following, we present detailed derivations of the optical conductivities for few-layer BP using the Kubo formula [Eq.~(\ref{eq:conductivity}) in the main text]. Note that for $M^{ss'}_i(\bm k)=\langle{s,\bm{k}}|\hbar\hat{v}_i |{s',\bm{k}'}\rangle$, $M^{ss'}_i(\bm k)M^{s's}_i(\bm k)=|M^{ss'}_i(\bm k)|^2$ ($i=x,y$) is always real. Thus, the intraband and interband contributions to the real part of the longitudinal conductivity in the clean limit are given by
\begin{align}
\sigma^{\mathrm{intra}}_{ii}(\omega)=- g_{\rm s} \frac{\pi e^2}{\hbar} \int \frac{d^2 k}{(2\pi)^2}
\sum_{s=\pm}
\frac{\partial f_{s,\bm{k}}}{\partial \varepsilon_{s,\bm{k}}} |M^{ss}_i(\bm k)|^2\delta(\hbar\omega)
\label{eq:kubo_intra}
\end{align}
and
\begin{align}
\sigma^{\rm inter}_{ii}(\omega)
&=
- g_{\rm s} \frac{\pi e^2}{\hbar} \int \frac{d^2 k}{(2\pi)^2}
\frac{f_{+,\bm{k}}-f_{-,\bm{k}}}{\varepsilon_{+,\bm{k}}-\varepsilon_{-,\bm{k}}} |M^{+-}_i(\bm k)|^2 \delta(\hbar\omega+\varepsilon_{-,\bm{k}}-\varepsilon_{+,\bm{k}})
\label{eq:kubo_inter}
\end{align}
at positive frequencies ($\omega>0$). Using this formula, it is straightforward to obtain the longitudinal optical conductivity for few-layer BP.

The model Hamiltonian for few-layer BP in Eq.~(\ref{eq:continuum_model_2}) in the main text can be simplified by introducing dimensionless parameters, and reduces to
\begin{align}
\label{eq:continuum_model_2_dimensionless}
H = \varepsilon_0 \left[\widetilde{k}_x \sigma_y + \left(\frac{1}{2} \widetilde{\varepsilon}_{\rm g} + \widetilde{k}_y^2  + \gamma \widetilde{k}_x^2 \right) \sigma_z \right],
\end{align}
where $\widetilde{k}_{x,y}=k_{x,y}/k_0$ are dimensionless momenta $k_0 = \frac{2m v}{\hbar}$, $\varepsilon_0 =  2m v^2$,  $\widetilde{\varepsilon}_{\rm g} = \varepsilon_{\rm g}/\varepsilon_0$, and the parameter $\gamma$ represents the ratio of the effective mass along the zigzag direction to that in the armchair direction. In the following, we simplify the continuum Hamiltonian [Eq.~(\ref{eq:continuum_model_2_dimensionless})] by setting $\gamma=0$, which allows an analytic form of the optical conductivity to be obtained.

\subsection{Intraband conductivity}
From Eq.~(\ref{eq:kubo_intra}), the intraband conductivity is expressed as
\begin{align}
\label{eq:conductivity_intra}
\sigma^{\mathrm{intra}}_{ii}(\omega)
&=
- g_{\rm s} \frac{\pi e^2}{\hbar} \int \frac{d^2 k}{(2\pi)^2}
\left(
\frac{\partial f_{+,\bm{k}}}{\partial \varepsilon_{+,\bm{k}}} |M^{++}_i(\bm k)|^2
+
\frac{\partial f_{-,\bm{k}}}{\partial \varepsilon_{-,\bm{k}}} |M^{--}_i(\bm k)|^2
\right)
\delta(\hbar\omega) \nonumber \\
&=
- g_{\rm s} \frac{\pi e^2}{\hbar} \int \frac{d^2 k}{(2\pi)^2}
\left(
\frac{\partial f_{+,\bm{k}}}{\partial \varepsilon_{+,\bm{k}}}
+
\frac{\partial f_{-,\bm{k}}}{\partial \varepsilon_{-,\bm{k}}}
\right)
|M_i^\mathrm{intra}(\bm k)|^2
\delta(\hbar\omega),
\end{align}
where the intraband matrix elements are given by
\begin{subequations}
\begin{eqnarray}
|M_x^\mathrm{intra}(\bm k)|^2&=&|M^{++}_x(\bm k)|^2 =|M^{--}_x(\bm k)|^2 = \frac{\varepsilon_0^2 \sin^2\psi}{k_0^2}, \\
|M_y^\mathrm{intra}(\bm k)|^2&=&|M^{++}_y(\bm k)|^2 =|M^{--}_y(\bm k)|^2 = \frac{4 \varepsilon_0^2 \widetilde{k}_y^2 \cos^2\psi}{k_0^2},
\end{eqnarray}
\end{subequations}
and $\psi = \tan^{-1}\left[\widetilde{k}_x/(\widetilde{\varepsilon}_{\rm g}/2 + \widetilde{k}_y^2)\right]$.

The intraband optical conductivity in Eq.~(\ref{eq:conductivity_intra}) can be rewritten as
\begin{subequations}
\begin{eqnarray}
\frac{\sigma^{\mathrm{intra}}_{xx}}{\sigma_0} &=& -g_{\rm s}\frac{1}{\pi} \int d^2\widetilde{k}
\left(\frac{\partial f_{+,\bm{k}}}{\partial \widetilde{\varepsilon}_{\bm k}}+\frac{\partial f_{-,\bm{k}}}{\partial (-\widetilde{\varepsilon}_{\bm k})}\right)
\frac{\widetilde{k}^2_x}{\widetilde{\varepsilon}^2_{\bm k}}
\delta\left({\hbar \omega \over \varepsilon_0}\right), \\
%%%%%%%%%%%%%
\frac{\sigma^{\mathrm{intra}}_{yy}}{\sigma_0} &=& -g_{\rm s}\frac{4}{\pi} \int d^2\widetilde{k}
\left(\frac{\partial f_{+,\bm{k}}}{\partial \widetilde{\varepsilon}_{\bm k}}+\frac{\partial f_{-,\bm{k}}}{\partial (-\widetilde{\varepsilon}_{\bm k})}\right)
\frac{\widetilde{k}^2_y \left(\widetilde{\varepsilon}_{\rm g}/2 + \widetilde{k}_y^2 \right)^2}{\widetilde{\varepsilon}^2_{\bm k}}
\delta\left({\hbar \omega \over \varepsilon_0}\right),
\end{eqnarray}
\end{subequations}
where $\sigma_0=e^2/(4\hbar)$, $\widetilde{\varepsilon}_{\bm k}=\varepsilon_+({\bm k})/\varepsilon_0=\sqrt{\left(\frac{1}{2} \widetilde{\varepsilon}_{\rm g} + \widetilde{k}_x^2 \right)^2+\widetilde{k}_y^2}$ and $g_{\rm s}=2$ is the spin degeneracy.

At zero temperature, $f_{\pm,\bm k}=\Theta\left(\mu-\varepsilon_{\pm}({\bm k})\right)$ and so $\displaystyle{\lim_{T \rightarrow 0}\left(-\frac{\partial f_{\pm,\bm{k}}}{\partial \varepsilon_{\pm,\bm{k}}}\right)=\delta\left(\mu-\varepsilon_{\pm}({\bm k})\right)}$. By using the relation $\delta(f(x)) = \sum\limits_i \frac{\delta(x-x_i)}{|f'(x_i)|}$, where $f(x_i)=0$, the intraband optical conductivity at zero temperature is given by
\begin{subequations}
\begin{eqnarray}
\frac{\sigma^{\mathrm{intra}}_{xx}}{\sigma_0}&=&
g_{\rm s}
\frac{1}{\pi}
\int_{-\infty}^{\infty}
\int_{-\infty}^{\infty}
d\widetilde{k}_x d\widetilde{k}_y
\left[\frac{\widetilde{k}_x}{\sqrt{\left(\frac{1}{2}\widetilde{\varepsilon}_{\rm g}+\widetilde{k}_y^2\right)^2+\widetilde{k}_x^2}}\right]^2
\delta\left(\left(\frac{|\mu|}{\varepsilon_0}\right)-\sqrt{\left(\frac{1}{2} \widetilde{\varepsilon}_{\rm g} + \widetilde{k}_y^2 \right)^2+\widetilde{k}_x^2}\right)
\delta\left({\hbar \omega \over \varepsilon_0}\right),\nonumber \\
&=&
g_{\rm s}
\frac{4}{\pi}
\frac{1}{(|\mu|/\varepsilon_0)}
\int d\widetilde{k}_y
\sqrt{\left(\frac{|\mu|}{\varepsilon_0}\right)^2-\left(\frac{1}{2}\widetilde{\varepsilon}_{\rm g}+\widetilde{k}_y^2\right)^2}
\delta\left({\hbar \omega \over \varepsilon_0}\right),\\
\frac{\sigma^{\mathrm{intra}}_{yy}}{\sigma_0}&=&
g_{\rm s}
\frac{4}{\pi}
\int_{-\infty}^{\infty}
\int_{-\infty}^{\infty}
d\widetilde{k}_x d\widetilde{k}_y
\left[\frac{\widetilde{k}_y \left(\frac{1}{2}\widetilde{\varepsilon}_{\rm g}+\widetilde{k}_y^2\right)}{\sqrt{\left(\frac{1}{2}\widetilde{\varepsilon}_{\rm g}+\widetilde{k}_y^2\right)^2+\widetilde{k}_x^2}}\right]^2
\delta\left(\left(\frac{|\mu|}{\varepsilon_0}\right)-\sqrt{\left(\frac{1}{2} \widetilde{\varepsilon}_{\rm g} + \widetilde{k}_y^2 \right)^2+\widetilde{k}_x^2}\right)
\delta\left({\hbar \omega \over \varepsilon_0}\right).\nonumber \\
&=&
g_{\rm s}
\frac{16}{\pi}
\frac{1}{(|\mu|/\varepsilon_0)}
\int d\widetilde{k}_y
\frac{\widetilde{k}_y^2\left(\frac{1}{2}\widetilde{\varepsilon}_{\rm g}+\widetilde{k}_y^2\right)^2}{\sqrt{\left(\frac{|\mu|}{\varepsilon_0}\right)^2-\left(\frac{1}{2}\widetilde{\varepsilon}_{\rm g}+\widetilde{k}_y^2\right)^2}}
\delta\left({\hbar \omega \over \varepsilon_0}\right),
\end{eqnarray}
\end{subequations}

First, consider the semi-Dirac case $(\varepsilon_{\rm g} = 0)$.
By performing the integration with $\varepsilon_{\rm g}=0$,
%the relation $\delta(f(x)) = \sum\limits_i \frac{\delta(x-x_i)}{|f'(x_i)|}$, where $f(x_i)=0$,
we obtain
\begin{subequations}
\begin{eqnarray}
\frac{\sigma^{\mathrm{intra}}_{xx}}{\sigma_0}&=&
\left[
\frac{4}{3\sqrt{\pi}} \frac{\Gamma(1/4)}{\Gamma(3/4)}
\right]
\left(\frac{|\mu|}{\varepsilon_0}\right)^{\frac{1}{2}}
\delta\left({\hbar \omega \over \varepsilon_0}\right),
\\
\frac{\sigma^{\mathrm{intra}}_{yy}}{\sigma_0}&=&
\left[
\frac{96}{5\sqrt{\pi}} \frac{\Gamma(3/4)}{\Gamma(1/4)}
\right]
\left(\frac{|\mu|}{\varepsilon_0}\right)^{\frac{3}{2}}
\delta\left({\hbar \omega \over \varepsilon_0}\right),
\end{eqnarray}
\end{subequations}
%where $\mathcal{A}_x$ and $\mathcal{A}_y$ are given by
%\begin{subequations}
%\begin{eqnarray}
%\mathcal{A}_{x} &=& g_{\rm s}\frac{4}{\pi} \int_0^1 dt \sqrt{1-t^4} = \frac{4}{3\sqrt{\pi}} \frac{\Gamma(1/4)}{\Gamma(3/4)} \approx 2.22567. \\
%\mathcal{A}_{y} &=& g_{\rm s}\frac{16}{\pi}\int_0^1 dt \frac{t^6}{\sqrt{1-t^4}} = \frac{96}{5\sqrt{\pi}} \frac{\Gamma(3/4)}{\Gamma(1/4)} \approx 3.66125,
%\end{eqnarray}
%\end{subequations}
Here, $\Gamma(z)=\int_0^{\infty}t^{z-1}e^{-t} dt$ is the gamma function.

%Similarly, for the insulator phase ($\varepsilon_{\rm g}>0$) we obtain
%\begin{subequations}
%\begin{eqnarray}
%\frac{\sigma^{\mathrm{intra}}_{xx}}{\sigma_0}&=& \frac{8\sqrt{2}}{3\pi}\left(\frac{\mu}{\varepsilon_0}\right)^{\frac{1}{2}} \left[ -2p {\rm E}\left(\frac{1-p}{2}\right) + (1+p) {\rm K}\left(\frac{1-p}{2}\right)\right] \Theta(\mu-\varepsilon_{\rm g}) \delta\left({\hbar \omega \over \varepsilon_0}\right), \\
%\frac{\sigma^{\mathrm{intra}}_{yy}}{\sigma_0}&=& \frac{32\sqrt{2}}{15\pi}\left(\frac{\mu}{\varepsilon_0}\right)^{\frac{3}{2}} \left[(18-4p^2) {\rm E}\left(\frac{1-p}{2}\right) + (1+p)(2p-9) {\rm K}\left(\frac{1-p}{2}\right)\right] \Theta(\mu-\varepsilon_{\rm g}) \delta\left({\hbar \omega \over \varepsilon_0}\right).
%\end{eqnarray}
%\end{subequations}
%Here, $p=\varepsilon_{\rm g}/2\mu$, $\mathrm{E}(x)$ and $\mathrm{K}(x)$ are the complete elliptic integral of the first and second kind, respectively.

For the insulator phase, the Drude weight defined to be $\sigma_{ii} = \mathcal{D}_{ii} \delta(\hbar \omega)$ in the vicinity of $\mu = \varepsilon_{\rm g}$, is given by
\begin{subequations}
\begin{eqnarray}
\frac{\mathcal{D}_{xx}}{\mathcal{D}_{0}} &\approx& \mathcal{B}_{x}^{(0)} + \mathcal{B}_{x}^{(1)} \left(\frac{\mu-\varepsilon_{\rm g}}{\varepsilon_0}\right) + O(\mu^2), \\
\frac{\mathcal{D}_{yy}}{\mathcal{D}_{0}} &\approx& \mathcal{B}_{y}^{(0)} + \mathcal{B}_{y}^{(1)} \left(\frac{\mu-\varepsilon_{\rm g}}{\varepsilon_0}\right) + O(\mu^2),
\end{eqnarray}
\end{subequations}
where $\mathcal{D}_{0}=\frac{e^2\varepsilon_0}{4\hbar}$, $\mathcal{B}_{x}^{(0)} = \frac{4\sqrt{2}}{3\pi} \left[-2{\rm E}({1\over 4})+3{\rm K}({1\over 4})\right]\sqrt{\widetilde{\varepsilon}_{\rm g} } \approx 1.274 \sqrt{\widetilde{\varepsilon}_{\rm g}}$, $\mathcal{B}_{x}^{(1)} = \frac{8\sqrt{2}}{3\pi} {\rm E}(\frac{1}{4})\frac{1}{\sqrt{\widetilde{\varepsilon}_{\rm g}}} \approx 1.762\frac{1}{\sqrt{\widetilde{\varepsilon}_{\rm g}}}$, $\mathcal{B}_{y}^{(0)} = \frac{32\sqrt{2}}{15\pi} \left[17{\rm E}({1\over 4})-12{\rm K}({1\over 4})\right] \widetilde{\varepsilon}_{\rm g}^{\frac{3}{2}} \approx 4.531 \widetilde{\varepsilon}_{\rm g}^{\frac{3}{2}}$, and $\mathcal{B}_{y}^{(1)} = \frac{64\sqrt{2}}{15\pi} \left[14{\rm E}({1\over 4})-9{\rm K}({1\over 4})\right] \widetilde{\varepsilon}_{\rm g}^{\frac{1}{2}} \approx 10.319 \widetilde{\varepsilon}_{\rm g}^{\frac{1}{2}}$.

For the Dirac semimetal phase ($\varepsilon_{\rm g}<0$),
%we find
%\begin{subequations}
%\begin{eqnarray}
%\frac{\sigma^{\mathrm{intra}}_{xx}}{\sigma_0}&=& {\rm FULL\ ANALYTIC\ FORM}, \\
%\frac{\sigma^{\mathrm{intra}}_{yy}}{\sigma_0}&=& {\rm FULL\ ANALYTIC\ FORM}.
%\end{eqnarray}
%\end{subequations}
the Drude weight at low Fermi energy is given by
\begin{subequations}
\begin{eqnarray}
\frac{\mathcal{D}_{xx}}{\mathcal{D}_{0}} &\approx& \mathcal{C}_{x}^{(0)} + \mathcal{C}_{x}^{(1)} \left(\frac{\mu-|\varepsilon_{\rm g}|}{\varepsilon_0}\right) + O(\mu^2), \\
\frac{\mathcal{D}_{yy}}{\mathcal{D}_{0}} &\approx& \mathcal{C}_{y}^{(0)} + \mathcal{C}_{y}^{(1)} \left(\frac{\mu-|\varepsilon_{\rm g}|}{\varepsilon_0}\right) + O(\mu^2),
\end{eqnarray}
\end{subequations}
where $\mathcal{C}_{x}^{(0)} = \frac{4\sqrt{2}}{3\pi} \left[2{\rm E}({3\over 4})+{\rm K}({3\over 4})\right]\sqrt{|\widetilde{\varepsilon}_{\rm g}|} \approx 2.748 \sqrt{|\widetilde{\varepsilon}_{\rm g}|}$, $\mathcal{C}_{x}^{(1)} = \frac{8\sqrt{2}}{3\pi} \left[-{\rm E}(\frac{3}{4})+{\rm K}(\frac{3}{4})\right]\frac{1}{\sqrt{|\widetilde{\varepsilon}_{\rm g}|}} \approx 1.762\frac{1}{\sqrt{|\widetilde{\varepsilon}_{\rm g}|}}$, $\mathcal{C}_{y}^{(0)} = \frac{32\sqrt{2}}{15\pi} \left[17{\rm E}({3\over 4})-5{\rm K}({3\over 4})\right] |\widetilde{\varepsilon}_{\rm g}|^{\frac{3}{2}} \approx 9.416 |\widetilde{\varepsilon}_{\rm g}|^{\frac{3}{2}}$, and $\mathcal{C}_{y}^{(1)} = \frac{64\sqrt{2}}{15\pi} \left[14{\rm E}({3\over 4})-5{\rm K}({3\over 4})\right] |\widetilde{\varepsilon}_{\rm g}|^{\frac{1}{2}} \approx 11.855 |\widetilde{\varepsilon}_{\rm g}|^{\frac{1}{2}}$.

At low densities (or Fermi energy), these analytic results are in good agreement with the results obtained using the lattice model (see Fig.~\ref{fig:optical_conductivity_intraband} in the main text). However, as the Fermi energy increases the analytic results deviate from the lattice results, especially in the Dirac semimetal phase, because the effect of the parabolic term $\gamma \frac{\hbar^2 k_x^2}{2m}$ in Eq.~(\ref{eq:continuum_model_2}) becomes significant.

%These analytic results are in good agreement at the conduction band minimum (or valence band maximum) with the result calculated based on the lattice model in Fig.~(\ref{fig:optical_conductivity_intraband}) in the main text. As the Fermi energy moves away from the conduction band minimum (or valence band maximum), the analytic result starts deviating from the lattice result due to the effect of the parabolic term $\gamma \frac{\hbar^2 k_x^2}{2m}$ in Eq.~(\ref{eq:continuum_model_2}) in the main text.

\subsection{Interband conductivity}
From Eq.~(\ref{eq:kubo_inter}), the interband conductivity is given by
\begin{eqnarray}
\sigma^{\mathrm{inter}}_{ii}(\omega)
&=&
- g_{\rm s} \frac{\pi e^2}{\hbar} \int \frac{d^2 k}{(2\pi)^2}
\frac{f_{+,\bm{k}}-f_{-,\bm{k}}}{\varepsilon_{+,\bm{k}}-\varepsilon_{-,\bm{k}}} |M^{+-}_i(\bm k)|^2 \delta(\hbar\omega+\varepsilon_{-,\bm{k}}-\varepsilon_{+,\bm{k}}).
\end{eqnarray}
In the presence of an electron-hole symmetry ($\varepsilon_{+,\bm{k}} = -\varepsilon_{-,\bm{k}}$), we can conveniently factor out the temperature dependence into a single coefficient:
\begin{align}
\sigma^{\mathrm{inter}}_{ii}(\omega)
\nonumber
&=
- g_{\rm s} \frac{\pi e^2}{\hbar} \int \frac{d^2 k}{(2\pi)^2}
\frac{f_{+,\bm{k}}-f_{-,\bm{k}}}{\varepsilon_{+,\bm{k}}-\varepsilon_{-,\bm{k}}} |M^{+-}_i(\bm k)|^2 \delta(\hbar\omega+\varepsilon_{-,\bm{k}}-\varepsilon_{+,\bm{k}}) \\
\nonumber
&=
- g_{\rm s} \frac{\pi e^2}{\hbar} \int \frac{d^2 k}{(2\pi)^2}
\frac{f\left(\varepsilon_{+,\bm{k}}\right)-f\left(-\varepsilon_{+,\bm{k}}\right)}{2\varepsilon_{+,\bm{k}}} |M^{+-}_i(\bm k)|^2 \delta(\hbar\omega-2\varepsilon_{+,\bm{k}}) \\
\nonumber
&=
\left[
f\left(-{\hbar \omega \over 2} \right)-f\left({\hbar \omega \over 2}\right)
\right]
\left[
g_{\rm s} \frac{\pi e^2}{\hbar} \int \frac{d^2 k}{(2\pi)^2}
\frac{|M^{+-}_i(\bm k)|^2}{{2\varepsilon_{+,\bm{k}}}} \delta(\hbar\omega-2\varepsilon_{+,\bm{k}})
\right] \\
&=
A(\omega, T, \mu) \ \sigma^{\mathrm{inter}}_{ii}(\omega, T=0, \mu=0),
\end{align}
where
\begin{align}
A(\omega, T, \mu)
=
f\left(-{\hbar \omega \over 2} \right)-f\left({\hbar \omega \over 2}\right)
=
\frac{\sinh{(\beta\hbar\omega/2)}}{\cosh{(\beta\hbar\omega/2)}+\cosh{(\beta\mu)}}.
\end{align}
Note that $A(\omega, T, \mu)$ absorbs all the dependence on the temperature and the chemical potential. In the following, we use $\sigma^{\mathrm{inter}}_{ii}(\omega)$ to refer to $\sigma^{\mathrm{inter}}_{ii}(\omega, T=0, \mu=0)$ for simplicity.
%In the presence of a symmetry between the conduction and valence bands, the interband conductivity can be factored into the product of the temperature factor and the interband conductivity at $T=0$ and $\mu=0$ case. Since $\varepsilon_{+,\bm{k}} = -\varepsilon_{-,\bm{k}}$, the interband conductivity reduces as follows:

With the interband matrix elements given by
\begin{subequations}
\begin{eqnarray}
|M_x^\mathrm{inter}(\bm k)|^2&=&M^{+-}_x(\bm k) M^{-+}_x(\bm k) = \frac{\varepsilon_0^2 \cos^2\psi}{k_0^2}, \\
|M_y^\mathrm{inter}(\bm k)|^2&=&M^{+-}_y(\bm k) M^{-+}_y(\bm k) = \frac{4 \varepsilon_0^2 \widetilde{k}_y^2 \sin^2\psi}{k_0^2},
\end{eqnarray}
\end{subequations}
where $\psi = \tan^{-1}\left[\widetilde{k}_x/(\widetilde{\varepsilon}_{\rm g}/2 + \widetilde{k}_y^2)\right]$, we can rewrite the interband conductivity as
\begin{align}
\nonumber
\frac{\sigma^{\mathrm{inter}}_{xx}}{\sigma_0}
&=
g_{\rm s} \frac{\pi e^2}{\hbar} \int \frac{d^2 k}{(2\pi)^2}
\frac{|M^{+-}_x(\bm k)|^2}{{2\varepsilon_{+,\bm{k}}}} \delta(\hbar\omega-2\varepsilon_{+,\bm{k}})
\\ \nonumber
&=
g_{\rm s}\frac{1}{2\pi} \int d^2\widetilde{k}
\frac{\left(\frac{1}{2}\widetilde{\varepsilon}_{\rm g}+\widetilde{k}_y^2\right)^2}{\widetilde{\varepsilon}^3_{\bm k}}
\delta \left( \widetilde{\omega}-2\widetilde{\varepsilon}_{\bm k} \right)
\\ \nonumber
&= g_{\rm s} \frac{2}{\pi}\int_{0}^{\infty}\int_{0}^{\infty} d\widetilde{k}_x d\widetilde{k}_y
\frac{\left(\frac{1}{2}\widetilde{\varepsilon}_{\rm g}+\widetilde{k}_y^2\right)^2}{\left[\sqrt{\left(\frac{1}{2} \widetilde{\varepsilon}_{\rm g} + \widetilde{k}_y^2 \right)^2+\widetilde{k}_x^2}\right]^3}
\delta
\left(
\widetilde{\omega}-2\sqrt{\left(\frac{1}{2} \widetilde{\varepsilon}_{\rm g} + \widetilde{k}_y^2 \right)^2+\widetilde{k}_x^2}
\right)
\\ \nonumber
&= g_{\rm s} \frac{1}{\pi}\int_{0}^{\infty}\int_{0}^{\infty} d\widetilde{k}_x d\widetilde{k}_y
\frac{\left(\frac{1}{2}\widetilde{\varepsilon}_{\rm g}+\widetilde{k}_y^2\right)^2}{(\widetilde{\omega}/2)^2 \widetilde{k}_x}
\delta
\left(
k_x - \sqrt{\left(\frac{\widetilde{\omega}}{2} \right)^2 - \left( \frac{1}{2} \widetilde{\varepsilon}_{\rm g}  + \widetilde{k}_y^2 \right )^2}
\right)
\\
&= g_{\rm s} \frac{1}{\pi{(\widetilde{\omega}/2)^2}}\int d\widetilde{k}_y
\frac{\left(\frac{1}{2}\widetilde{\varepsilon}_{\rm g}+\widetilde{k}_y^2\right)^2}{\sqrt{\left(\frac{\widetilde{\omega}}{2} \right)^2 - \left( \frac{1}{2} \widetilde{\varepsilon}_{\rm g}  + \widetilde{k}_y^2 \right )^2}},
\label{eq:sigmaxx}
%%%%%%%%%%%%%%%%
\\ \nonumber
\frac{\sigma^{\mathrm{inter}}_{yy}}{\sigma_0}
&=
g_{\rm s} \frac{\pi e^2}{\hbar} \int \frac{d^2 k}{(2\pi)^2}
\frac{|M^{+-}_y(\bm k)|^2}{{2\varepsilon_{+,\bm{k}}}} \delta(\hbar\omega-2\varepsilon_{+,\bm{k}})
\\ \nonumber
&=
g_{\rm s}\frac{2}{\pi} \int d^2\widetilde{k}
\frac{\widetilde{k}_x^2 \widetilde{k}_y^2}{\widetilde{\varepsilon}^3_{\bm k}}
\delta \left( \widetilde{\omega}-2\widetilde{\varepsilon}_{\bm k} \right)
\\ \nonumber
&= g_{\rm s} \frac{8}{\pi}\int_{0}^{\infty}\int_{0}^{\infty} d\widetilde{k}_x d\widetilde{k}_y
\frac{\widetilde{k}_x^2 \widetilde{k}_y^2}{\left[\sqrt{\left(\frac{1}{2} \widetilde{\varepsilon}_{\rm g} + \widetilde{k}_y^2 \right)^2+\widetilde{k}_x^2}\right]^3}
\delta
\left(
\widetilde{\omega}-2\sqrt{\left(\frac{1}{2} \widetilde{\varepsilon}_{\rm g} + \widetilde{k}_y^2 \right)^2+\widetilde{k}_x^2}
\right)
\\ \nonumber
&= g_{\rm s} \frac{4}{\pi}\int_{0}^{\infty}\int_{0}^{\infty} d\widetilde{k}_x d\widetilde{k}_y
\frac{\widetilde{k}_x \widetilde{k}_y^2}{(\widetilde{\omega}/2)^2}
\delta
\left(
k_x - \sqrt{\left(\frac{\widetilde{\omega}}{2} \right)^2 - \left( \frac{1}{2} \widetilde{\varepsilon}_{\rm g}  + \widetilde{k}_y^2 \right )^2}
\right)
\\
&= g_{\rm s} \frac{4}{\pi{(\widetilde{\omega}/2)^2}}\int d\widetilde{k}_y
\widetilde{k}_y^2
\sqrt{\left(\frac{\widetilde{\omega}}{2} \right)^2 - \left( \frac{1}{2} \widetilde{\varepsilon}_{\rm g}  + \widetilde{k}_y^2 \right )^2},
\label{eq:sigmayy}
\end{align}
where $\sigma_0=e^2/(4\hbar)$, $\widetilde{\omega}=\hbar \omega /\varepsilon_0$, $\widetilde{\varepsilon}_{\bm k}=\varepsilon_+({\bm k})/\varepsilon_0=\sqrt{\left(\frac{1}{2} \varepsilon_g + \widetilde{k}_x^2 \right)^2+\widetilde{k}_y^2}$, $f(\widetilde{\varepsilon}_{\bm k})=1/[1+e^{\beta(\varepsilon_0\widetilde{\varepsilon}_{\bm k}-\mu)}]$ and $g_{\rm s}=2$ is the spin degeneracy.

%Introducing a new variable $u = (\frac{1}{2}\widetilde{\varepsilon}_{\rm g}+\widetilde{k}_x^2)/(\frac{\widetilde{\omega}}{2})$, Eqs.~(\ref{eq:sigmaxx}) and (\ref{eq:sigmayy}) become
%\begin{subequations}
%\label{eq:sigma_interband_analytic}
%\begin{eqnarray}
%\frac{\sigma^{\mathrm{inter}}_{xx}}{\sigma_0} &=&
%g_{\rm s} \frac{\sqrt{2\widetilde{\omega}}}{\pi}
%\int du
%\sqrt{u-p}\sqrt{1-u^2},\\
%\frac{\sigma^{\mathrm{inter}}_{yy}}{\sigma_0} &=&
%\frac{g_{\rm s}}{\pi\sqrt{2\widetilde{\omega}}}
%\int du
%\frac{u^2}{\sqrt{u-p}\sqrt{1-u^2}},
%\end{eqnarray}
%\end{subequations}
%where $p\equiv \varepsilon_{\rm g}/(\hbar\omega)$. Here, the integration range depends on the sign of $\varepsilon_{\rm g}$, which is given in Tab.~A1. By evaluating the above integrals with $g_{\rm s}=2$, we obtain the analytic expressions of the optical conductivities for each phase.
%
%\setcounter{table}{0}
%\renewcommand{\thetable}{A\arabic{table}}
%
%\begin{table}[htp]
%\label{tab:integration_range}
%\begin{tabular}{c |c| c}
% &$\hbar\omega<|\varepsilon_{\rm g}|$ & $\hbar\omega>|\varepsilon_{\rm g}|$ \\
%\hline
%$\varepsilon_{\rm g}>0$ & vanishes & $[p,1]$ \\
%$\varepsilon_{\rm g}=0$ &- & $[0,1]$ \\
%$\varepsilon_{\rm g}<0$ & $[-1,1]$ & $[p,1]$
%\end{tabular}
%\caption{Integration range in Eq.~(\ref{eq:sigma_interband_analytic}) depending on the sign of $\varepsilon_{\rm g}$.}
%\end{table}
%%Now, consider the interband optical conductivity for each phase at zero temperature ($T=0$) for the undoped case ($\mu=0$).

Now, consider the undoped case ($\mu=0$).
For the semi-Dirac point ($\varepsilon_{\rm g}=0$), we obtain
\begin{subequations}
\begin{eqnarray}
\frac{\sigma^{\mathrm{inter}}_{xx}(\omega)}{\sigma_0}  &=&
\left[\frac{1}{3\sqrt{2\pi}} \frac{\Gamma(1/4)}{\Gamma(3/4)} \right] \left(\frac{\omega}{\omega_0}\right)^{-{1\over 2}},\\
\frac{\sigma^{\mathrm{inter}}_{yy}(\omega)}{\sigma_0} &=&
\left[\frac{8\sqrt{2}}{5\sqrt{\pi}} \frac{\Gamma(3/4)}{\Gamma(1/4)} \right] \left(\frac{\omega}{\omega_0}\right)^{{1\over 2}}.
\end{eqnarray}
\end{subequations}

%\subsubsection{Insulator phase ($\varepsilon_{\rm g}>0$)}
For the insulator phase ($\varepsilon_{\rm g}>0$), we have
\begin{subequations}
\begin{eqnarray}
\frac{\sigma^{\mathrm{inter}}_{xx}(\omega)}{\sigma_0} &=&
\frac{2}{3\pi}
\frac{1}{\sqrt{\widetilde{\omega}}}
\left[
4p
{\rm E} \left(\frac{1-p}{2} \right)
+(1-2p)
{\rm K} \left(\frac{1-p}{2} \right)
\right]
\Theta(\hbar\omega-\varepsilon_{\rm g}),
\\
\frac{\sigma^{\mathrm{inter}}_{yy}(\omega)}{\sigma_0} &=&
\frac{8}{15\pi}
\sqrt{\widetilde{\omega}}
\left[
2(3+p^2)
{\rm E} \left(\frac{1-p}{2} \right)
-(1+p)(3+p)
{\rm K}\left(\frac{1-p}{2} \right)
\right]
\Theta(\hbar\omega-\varepsilon_{\rm g}),
\end{eqnarray}
\end{subequations}
where $p = \varepsilon_{\rm g} /(\hbar\omega)$. Note that the integral vanishes when $\hbar\omega<\varepsilon_{\rm g}$, representing an optical gap.

%\subsubsection{Semi-metallic phase ($\varepsilon_{\rm g}<0$)}
For the Dirac semimetal phase ($\varepsilon_{\rm g}<0$), we find
\begin{subequations}
\begin{eqnarray}
\frac{\sigma^{\mathrm{inter}}_{xx}(\omega)}{\sigma_0} \bigg\rvert_{\hbar \omega <|\varepsilon_{\rm g}|} &=&
\frac{2\sqrt{2}}{3\pi}\frac{1}{\sqrt{\widetilde{\omega}}}
\frac{1}{\sqrt{1-p}}
\left[
2p(1-p)
{\rm E} \left(\frac{2}{1-p} \right)
+
(1+2p^2)
{\rm K} \left(\frac{2}{1-p} \right)
\right],
\\
\frac{\sigma^{\mathrm{inter}}_{xx}(\omega)}{\sigma_0}  \bigg\rvert_{\hbar \omega >|\varepsilon_{\rm g}|} &=&
\frac{2}{3\pi}
\frac{1}{\sqrt{\widetilde{\omega}}}
\left[
4p
{\rm E} \left(\frac{1-p}{2} \right)
+(1-2p)
{\rm K} \left(\frac{1-p}{2} \right)
\right].
\\
\frac{\sigma^{\mathrm{inter}}_{yy}(\omega)}{\sigma_0} \bigg\rvert_{\hbar \omega <|\varepsilon_{\rm g}|} &=&
\frac{8\sqrt{2}}{15\pi} \sqrt{\widetilde{\omega}}
\sqrt{1-p}
\left[
(3+p^2)
{\rm E} \left(\frac{2}{1-p} \right)
-p(1+p)
{\rm K} \left(\frac{2}{1-p} \right)
\right],
\\
\frac{\sigma^{\mathrm{inter}}_{yy}(\omega)}{\sigma_0} \bigg\rvert_{\hbar \omega >|\varepsilon_{\rm g}|} &=&
\frac{8}{15\pi}
\sqrt{\widetilde{\omega}}
\left[
2(3+p^2)
{\rm E} \left(\frac{1-p}{2} \right)
-(1+p)(3+p)
{\rm K} \left(\frac{1-p}{2} \right)
\right],
\end{eqnarray}
\end{subequations}

The effect of non-zero $T$ and $\mu$ on the optical conductivity can be taken into account by multiplying the temperature factor $A(\omega, T, \mu)$ into the $T=0$ and $\mu=0$ case.

\section{Evolution of the optical peaks between the insulator phase and the semi-Dirac point}
\label{app:evolution}
In this section, we describe the evolution of the interband optical conductivity along the armchair direction ($\sigma_{xx}$) for tetralayer BP from the insulator phase to the semi-Dirac point, as the external electric field increases. Figure \ref{fig:optical_conductivity_insulator_evolution} shows $\sigma_{xx}$ for tetralayer BP for several values of $E_{\rm ext}$, and the evolution of the optical peaks as a function of $E_{\rm ext}$, between the insulator phase and the semi-Dirac point.

%%%%%%%%%%%%%%%%%%%%%%%%%%%%%%%%%%%%%%%%%%%%%%%%%%
\begin{figure}[htb]
\includegraphics[width=0.9\linewidth]{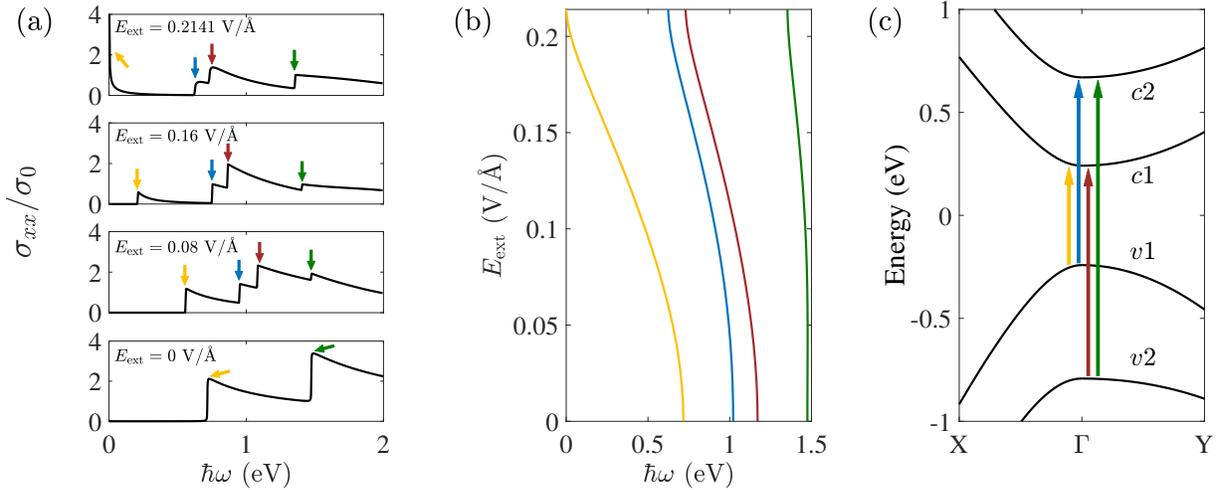}
\caption{
(a) Optical conductivity $\sigma_{xx}$ of tetralayer BP for $E_{\rm ext}=0$, $0.08$, $0.16$, $0.2141$ ${\rm V/\AA}$. (b) Evolution of the first four optical peaks in $\sigma_{xx}$ for tetralayer BP between the insulator phase and the semi-Dirac point. (c) The band structure of tetralayer BP in the insulator phase with $E_{\rm ext} = 0$ ${\rm V/\AA}$. Arrows indicate the interband transitions corresponding to the first four optical peaks in $\sigma_{xx}$. Here, $\sigma_0={e^2\over 4\hbar}$, and the lattice model with $\mu=0$ was utilized for the calculation.
}
\label{fig:optical_conductivity_insulator_evolution}
\end{figure}
%%%%%%%%%%%%%%%%%%%%%%%%%%%%%%%%%%%%%%%%%%%%%%%%%%

Note that the optical peaks only appear in $\sigma_{xx}$, but they are suppressed in $\sigma_{yy}$ along the zigzag direction, owing to the selection rule. As shown in Fig.~\ref{fig:atomic_structure} in the main text, few-layer BP has reflection symmetry with respect to the $y=0$ plane [$\mathcal{M}_y: (x,y) \rightarrow (x,-y) $]. According to the density functional theory calculation \cite{rudenko2014quasiparticle, rodin2014strain}, the conduction and valence bands at the $\Gamma$ point are composed of just the $3s$, $3p_x$, and $3p_z$ orbitals, which are all even in $\mathcal{M}_y$ ($\mathcal{M}_y\ket{3s, 3p_x, 3p_z} = +\ket{3s, 3p_x, 3p_z}$).
Therefore, the matrix element $M_y^{ss'}(\bm{k})$  [Eq.~(\ref{eq:conductivity}) in the main text] at the $\Gamma$ point can be expressed as $M_y^{ss'}(0)=
\braket{s,0|\frac{\partial H}{\partial k_y}|s',0} = \braket{s,0|\mathcal{M}_y^{\dagger}\mathcal{M}_y\frac{\partial H}{\partial k_y}\mathcal{M}_y^{\dagger}\mathcal{M}_y|s',0} = -\braket{s,0|\frac{\partial H}{\partial k_y}|s',0}$, leading to $M_y^{ss'}(0) = 0$, and thus suppressing $\sigma_{yy}$.

\end{widetext}
%%%%%%%%%%%%%%%%%%%%%%%%%%%%%%%%%%%%%%%%%%%%%%%%%%
%\bibliographystyle{apsrev4-1.bst}
\bibliographystyle{apscustom.bst}
\bibliography{bibfile}
%%%%%%%%%%%%%%%%%%%%%%%%%%%%%%%%%%%%%%%%%%%%%%%%%%
\end{document}